\documentclass[]{aa} 

\usepackage{relsize}
\usepackage{upgreek}
\usepackage{amsmath}

\usepackage{graphicx}
\usepackage{txfonts}
\begin{document}

   \title{Evidence for azimuthal variations of the oxygen abundance gradient tracing the spiral structure of the galaxy HCG~91c }
   \titlerunning{The azimuthal variations of the oxygen abundance gradient in HCG~91c}

   \author{F.~P.~A. Vogt\inst{1}\thanks{ESO Fellow} 
         \and
         E. P\'erez\inst{2}
         \and
         M.~A. Dopita\inst{3}
         \and
         L. Verdes-Montenegro\inst{2}
         \and
         S. Borthakur\inst{4} 
          }

   \institute{$^1$ European Southern Observatory, Av. Alonso de C\'ordova 3107, 763 0355 Vitacura, Santiago, Chile. \\  
                  $^2$ Instituto de Astrof\'isica de Andaluc\'ia, CSIC, Apdo. Correos 3004, E-18080 Granada, Spain.\\
                  $^3$ Research School of Astronomy and Astrophysics, Australian National University, Canberra, Australia.\\
                  $^4$ Department of Physics and Astronomy, Johns Hopkins University, 3400 N. Charles Street, Baltimore, MD 21218, USA.\\
      \email{frederic.vogt@alumni.anu.edu.au}
}

   \date{Received October 5, 2016; accepted December 23, 2016}
 
  \abstract
   {The distribution of elements in galaxies forms an important diagnostic tool to characterize the system's formation and evolution. This tool is however complex to use in practice, as galaxies are subject to a range of simultaneous physical processes active from pc to kpc scales. This renders observations of the full optical extent of galaxies down to sub-kpc scales essential.} 
   {Using the WiFeS integral field spectrograph, we previously detected abrupt and localized variations in the gas-phase oxygen abundance of the spiral galaxy HCG~91c. Here, we follow-up on these observations to map HCG~91c's disk out to $\sim$2\,R$_{e}$ at a resolution of 600\,pc, and characterize the non-radial variations of the gas-phase oxygen abundance in the system.}
   {We obtained deep MUSE observations of the target under $\sim$0.6 arcsec seeing conditions. We perform both a spaxel-based and aperture-based analysis of the data to map the spatial variations of 12+$\log$(O/H) across the disk of the galaxy.}
   {We confirm the presence of rapid variations of the oxygen abundance across the entire extent of the galaxy previously detected with WiFeS, for all azimuths and radii. The variations can be separated in two categories: a) \emph{localized} and associated with individual H\,\textsc{\smaller II} regions, and b) \emph{extended} over kpc scales, and occurring at the boundaries of the spiral structures in the galaxy.}
   {Our MUSE observations suggest that the enrichment of the interstellar medium in HGC~91c has proceeded preferentially along spiral structures, and less efficiently across them. Our dataset highlights the importance of distinguishing individual star-forming regions down to scales of a few 100\,pc when using integral field spectrographs to spatially resolve the distribution of oxygen abundances in a given system, and accurately characterize azimuthal variations and intrinsic scatter.}


   \maketitle
%

\section{Introduction}
The distribution within galaxies of the atomic elements created via stellar evolution is a key observable affected by the various processes associated with galaxy evolution. In principle, the physical mechanisms associated with specific processes affecting galaxies can be deciphered from a spatially-resolved characterization of the distribution of atomic elements heavier than Helium (located either in the gas phase or locked in subsequent generations of stars), provided that the spatial and temporal enrichment of the interstellar medium (ISM) in the system can be constrained. In practice, the co-existence of multiple evolutionary mechanisms effective on different spatial and temporal scales significantly complicates such an analysis for any given system. Evolutionary mechanisms include galactic gaseous outflows, gaseous inflows, stellar and gaseous radial migration (e.g. through the presence of a bar), gravitational interactions, mergers, accretion events and other forms of perturbations, including strangulation, harassment, and stripping.

In recent years, integral field spectrographs (IFSs) have largely supplanted long-slit spectrographs in studies designed to characterize the abundance distribution of heavy elements in galaxies. Among other benefits, the ability of IFS to measure abundances throughout the full two-dimensional extent of a galaxy (or a large part thereof) and detect azimuthal and radial trends has often been praised. In practice, the relatively small field-of-views (FoVs) and/or large size (on-sky) of the spatial pixels (a.k.a spaxels) of many IFSs have so far restricted the feasibility of performing such combined analysis of both sub-kpc trends with larger kpc-scales variations in systems located beyond a few Mpc. Instead, gaseous abundance distributions are often characterized via the slope of the azimuth-averaged gradient; an approach usually driven by the lack of spatial resolution and/or signal-to-noise (S/N), but that also allows for a more straight-forward comparison between galaxies at higher redshift, modulo the issues inherent to comparing oxygen abundances derived using different methodologies \citep[e.g.][]{Kewley2008, Lopez-Sanchez2012}.

The advent of the MUSE (Multi-Unit Spectroscopic Explorer) IFS has opened a new observational window as the combination of a relatively large FoV (1$\times$1\,square arcmin) and small spaxels (0.2$\times$0.2\,square arcsec) coupled to the excellent seeing of Cerro Paranal is now allowing to map galaxies out to 2-3 effective radius (R$_{e}$) and (simultaneously) down to sub-kpc scales out to distance of $\sim$100\,Mpc. MUSE is effectively pushing outwards our ability to study galaxies as complex systems down to the crucial sub-kpc scales (the scale of individual H\,{\smaller II} region complexes): a feat previously restricted to galaxies closer-in via IFS with larger FoV such as PMAS/PPAK \citep[e.g. via the PINGS and CALIFA surveys;][]{Rosales-Ortega2010,Sanchez2012,Sanchez2012a,Sanchez2014} and VIRUS-P \citep[e.g. via the VENGA survey; ][]{Blanc2013, Kaplan2016}. This push is particularly interesting from a galaxy evolution perspective, as MUSE effectively allows the study of galaxies inside a wider range of environments, including that of compact groups and clusters, with unprecedented spatial resolution.

In this article, we present follow-up MUSE observations of the nearly face-on spiral galaxy HCG~91c. Initial observations of this galaxy with the Wide Field Spectrograph \citep[WiFeS;][]{Dopita2007,Dopita2010} mounted on the ANU 2.3m telescope \citep{Mathewson2013} at Siding Spring observatory revealed spatially rapid and localized variations of the oxygen abundance in the system associated with at least one star-forming complex. HCG~91c is a member of a compact group of galaxies, which could indicate a possible influence of the environment on the chemical content within the galaxy. However, the WiFeS data was insufficient to distinguish between the possibility that gas of different metallicity had fallen in from another galaxy (or from the Intergalactic Medium), or whether the variation was caused by secular processes. The new MUSE data presented in Sec.~\ref{sec:obs} (thanks to the large FoV and 3$\times$ better spatial resolution of this instrument) now allows us to do this. Our analysis procedure is described in Sec.~\ref{sec:analysis}. We present our results in Sec.~\ref{sec:results} and discuss them in Sec.~\ref{sec:discussion}. Our conclusions are summarized in Sec.~\ref{sec:conclusion}.

\section{Observations and data reduction}\label{sec:obs}
HCG~91c was observed during the second Science Verification run for MUSE (mounted at the Nasmyth B focus of the Unit 4 -Yepun of the Very Large Telescope on Cerra Paranal), under program ID 60.A-9317(A) (P.I.: Vogt). The observation strategy for this program is described in detail in \cite{Vogt-thesis}, and summarized here for completeness. A total of 12 individual exposures (eleven of 1050\,s and one of 525\,s) on-target separated in 6 observation blocks (OBs) were acquired over 4 distinct nights between 2014, August 20 and 2014, August 25. In each OB, two exposures on-target were surrounding the observation of a dedicated empty sky field located near-by. Each on-target exposure was reduced individually using the \textsc{reflex} \citep{Freudling2013} MUSE workflow (v1.6), including the construction of the individual datacubes. Out of the 12 on-target exposures, 3 were acquired under seeing conditions $>$0.8\,arcsec, measured from the reconstructed individual datacubes using a star in the FoV (and consistent with the reported values of the Slow Guiding System). The other 9 exposures were acquired under seeing conditions $<$ 0.7 arcsec: these 9 exposures were combined together to form the final data cube presented in this article via the dedicated \textsc{reflex} \textsc{muse\_exp\_combine.xml} workflow. The combined datacube thus corresponds to $8\cdot1050+1\cdot525=8925$\,s on-source, and the spatial full-width at half-maximum (FWHM) of stars in the FoV are measured to be 0.6\,arcsec in the V-band. Spectrally, the cube extends from 4750\,\AA\ to 9350\,\AA\ in steps of 1.25\,\AA\footnote{The MUSE spectral resolution increases from $R\cong1700$ to $R\cong3500$ between 4750\,\AA\ and 9350\,\AA.}, with $319\times316$ spaxels. The individual exposures were acquired at four distinct position angles (P.A.; 0$^{\circ}$, 90$^{\circ}$, 180$^{\circ}$, 270$^{\circ}$) and with sub-arcsec spatial offsets to best remove the background-level artefacts associated with the 24 individual integral field units inside MUSE. The reduced datacube was uploaded to the ESO science archive facility following the recommendations of the \textit{Phase 3} stage of observations with ESO facilities\footnote{\url{https://www.eso.org/sci/observing/phase3.html}}. 

\begin{figure}
\centerline{\includegraphics[width=\hsize]{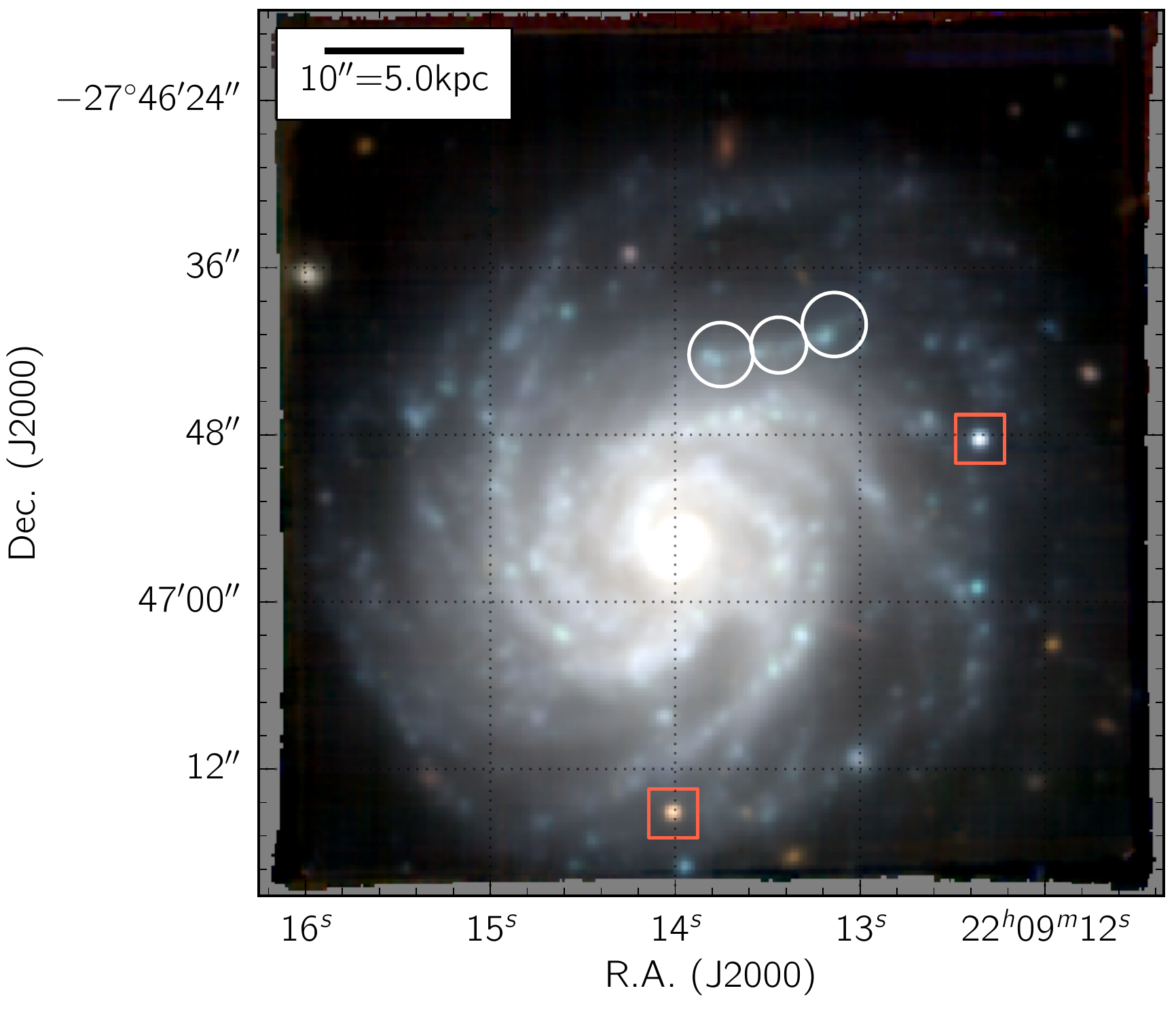}}
\caption{Pseudo-RGB image of HCG~91c, where the R, G \& B colors correspond to the sum of the reduced MUSE cube slices in the spectral ranges [7500\,\AA; 9300\,\AA], [6000\,\AA; 7500\,\AA] \& [4800\,\AA; 6000\,\AA], respectively. The spatial FWHM of the datacube is $\sim$0.6\,arcsec in the V-band. The star-forming complexes found by \cite{Vogt2015}  (then dubbed ``C1'', ``C2'' and ``C3'') to have discrepant oxygen abundances with respect to their immediate surroundings are marked with white circles. Two foreground stars (indicative of the spatial resolution of the data) are marked with red boxes.}
 \label{fig:RGB}
 \end{figure}

A pseudo-RGB image of the final datacube is presented in Figure~\ref{fig:RGB}, where the R, G \& B-bands correspond to the sum of the cube slices in the spectral ranges [7500\,\AA; 9300\,\AA], [6000\,\AA; 7500\,\AA] \& [4800\,\AA; 6000\,\AA], respectively. The spiral structure of the galaxy is readily evident, extending throughout the entire optical disk. Two foreground stars are visible at [22$^{h}$09$^{m}$14.\!\!$^{s}$02; -27$^{\circ}$47$^{\prime}$15.\!\!$^{\prime\prime}$1] and [22$^{h}$9$^{m}$12.\!\!$^{s}$37;  -27$^{\circ}$46$^{\prime}$48.\!\!$^{\prime\prime}$4], and provide a visual scale for the spatial resolution of the image. Numerous background galaxies are also visible as redder sources across the FoV. At the distance of HCG~91c \citep[104\,Mpc, see Table~\ref{table:HCG91c} and][]{Vogt2015}, 10\,arcsec correspond to 5\,kpc.

\begin{table}
{\smaller
\caption{Basic characteristics of HCG~91c.}\label{table:HCG91c}
\flushleft{
\begin{tabular}{l p{2.2cm} l}
\hline
\hline
Property & Value & Reference\\
\hline
Names & HCG~91c &\\
& ESO 467 - G 013 & \\[0.5ex]
R.A. [J2000] & 22$^{h}$09$^{m}$07.7$^{s}$ &\\[0.5ex]
Dec. [J2000] & -27$^{\circ}$48$^{\prime}$34$^{\prime\prime}$  & \\[0.5ex]
Redshift & 0.024377 &  \cite{Vogt2015}\\[0.5ex]
Distance & 104 Mpc &\\[0.5ex]
Spatial scale & 504 pc arcsec$^{-1}$ &\\[0.5ex]
R$_{25}$ & 26.75$\pm$3.25 arcsec &  \cite{deVaucouleurs1991}\\[0.5ex]
Rotation velocity & 100$\pm$11 km s$^{-1}$ &  \cite{Vogt2015}\\
\quad(at radii$>$22~kpc) & \\[0.5ex]
Star formation rate  & 2.19 M$_{\odot}$ yr$^{-1}$ & \cite{Bitsakis2014}\\
                                    & 2.10$\pm$0.06 M$_{\odot}$ yr$^{-1}$ &  \cite{Vogt2015}\\[0.5ex]
Stellar mass & 1.86$\times$10$^{10}$ M$_{\odot}$ & \cite{Bitsakis2014} \\[0.5ex]
H\,\textsc{\smaller I} mass  & 2.3$\times10^{10}$ M$_{\odot}$ & \cite{Borthakur2010}\\
\hline
\end{tabular}
}}
\end{table}

\section{Data post-processing}\label{sec:analysis}

With $>10^{5}$ spectra in the final MUSE datacube, any manual data processing step becomes extremely costly time-wise. Each second invested for the analysis of a single spectra would immediately require $\sim$28\,hours to manually perform the same task for all the spaxels in the MUSE datacube (one after another). To circumvent this time sink (also associated with the processing of large number of IFS observations from other instruments), several tools have been developed to process IFS data products in an automated fashion: recent examples include \textsc{pycasso} \citep{CidFernandes2013}, \textsc{lzifu} \citep[][]{Ho2016}, and \textsc{pipe3d} \citep[][]{Sanchez2016,Sanchez2016a}. For our analysis, we have developed our own post-processing tool called \textsc{brutus}: a set of \textsc{python} modules designed to automatically process datacubes from integral field spectrographs. \textsc{brutus} is hosted on Github, and is made freely available to the community\footnote{\url{http://fpavogt.github.io/brutus}}. It is designed with a modular structure in mind \citep[inspired by \textsc{pywifes};][]{Childress2014,Childress2014a}, allowing users to choose which processing steps are to be run (or not). A detailed description of the code is outside the scope of this article, but for completeness we list in Appendix~\ref{app:brutus} the specific steps employed to process the datacube of HCG~91c. 

\begin{figure}
\centerline{\includegraphics[width=\hsize]{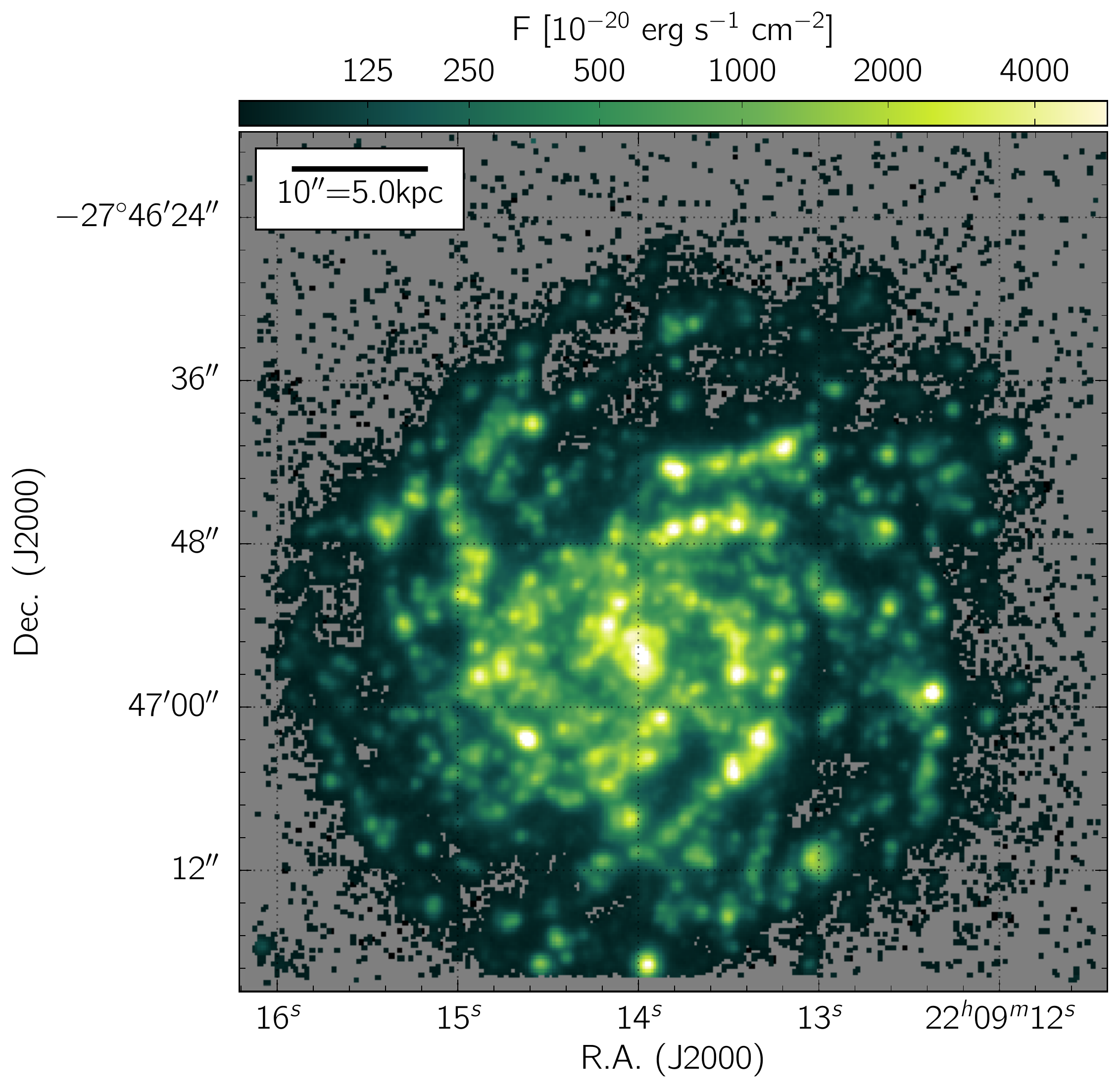}}
\centerline{\includegraphics[width=\hsize]{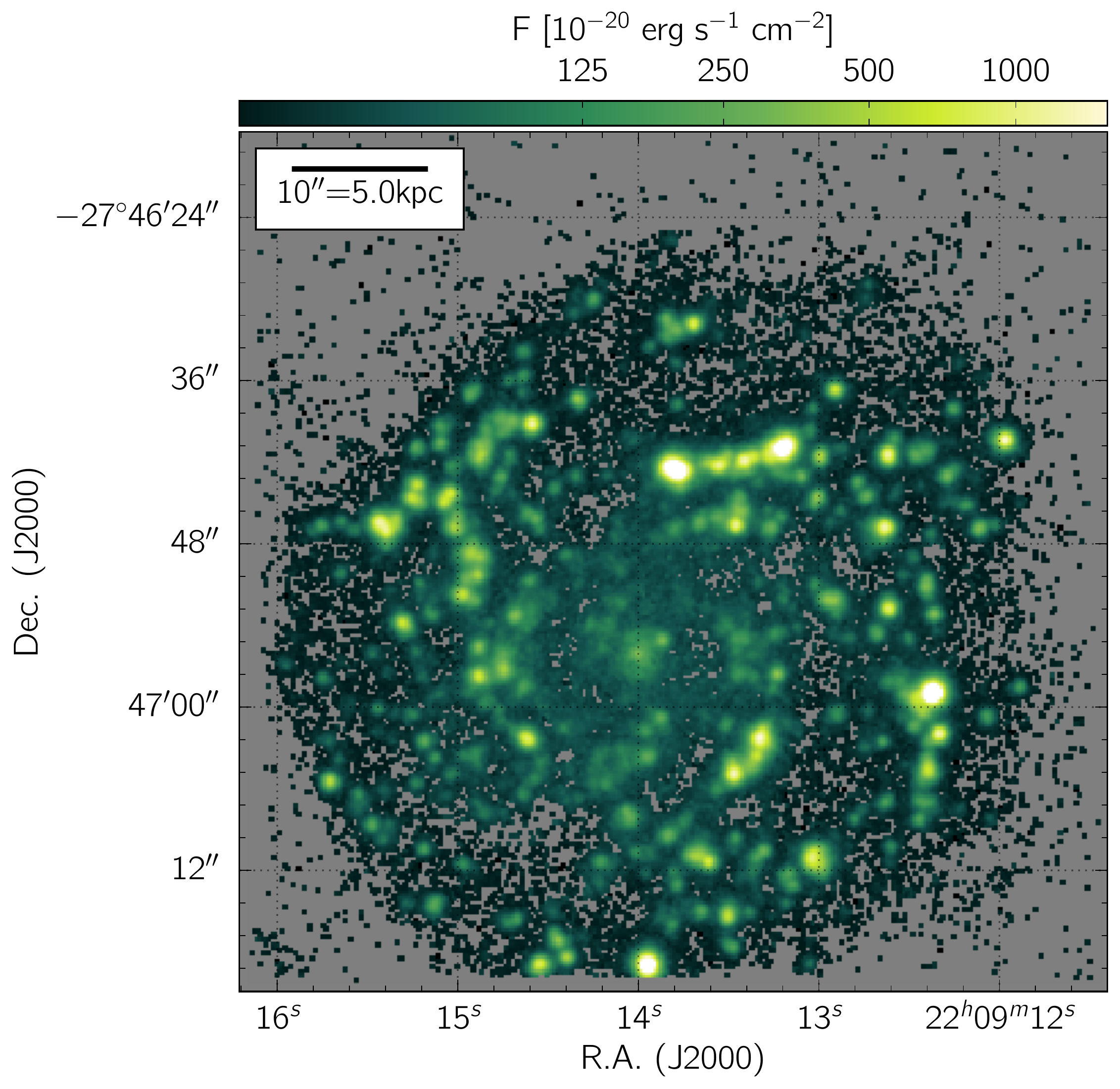}}
\caption{H$\alpha$ (top) and [O\,\textsc{\smaller III}]$ \lambda$5007 flux map of HCG~91c. The emission lines were fitted with tied gaussian components in each spaxel using \textsc{brutus}. For the H$\alpha$ map, all spaxels with SNR$\geq$5 are shown. For the [O\,\textsc{\smaller III}]$ \lambda$5007 map, all spaxels with SNR$\geq$3 are shown.}
 \label{fig:flux-maps}
 \end{figure}

The emission line flux maps for H$\alpha$ and [O\,\textsc{\smaller III}]$\lambda$5007 constructed using \textsc{brutus} are presented in Figure~\ref{fig:flux-maps}. In this work, we restrict our analysis to spaxels with SNR(H$\alpha$, H$\beta$)$\geq$5 and SNR$\geq$3 for all other lines: a good detection of the first two Hydrogen Balmer lines ensures reliable measurements of the tied velocity and velocity dispersions, hence leading to stable fits for the other lines for SNRs$\geq$3. A significant detection of the first two Hydrogen Balmer lines is also essential to ensure a reliable correction of the extragalactic attenuation on a spaxel-by-spaxel basis. In the present case, the extragalactic attenuation is corrected with \textsc{brutus} via the $H\alpha$ to H$\beta$ line flux ratio and the theoretical model of a turbulent dust screen from \cite{Fischera2005}, with $R_V=3.08$ and $R_V^A=4.3$ which in practice results in a correction function very similar to that of \cite{Calzetti2000} across the MUSE spectral range.

\begin{figure}
\centerline{\includegraphics[width=\hsize]{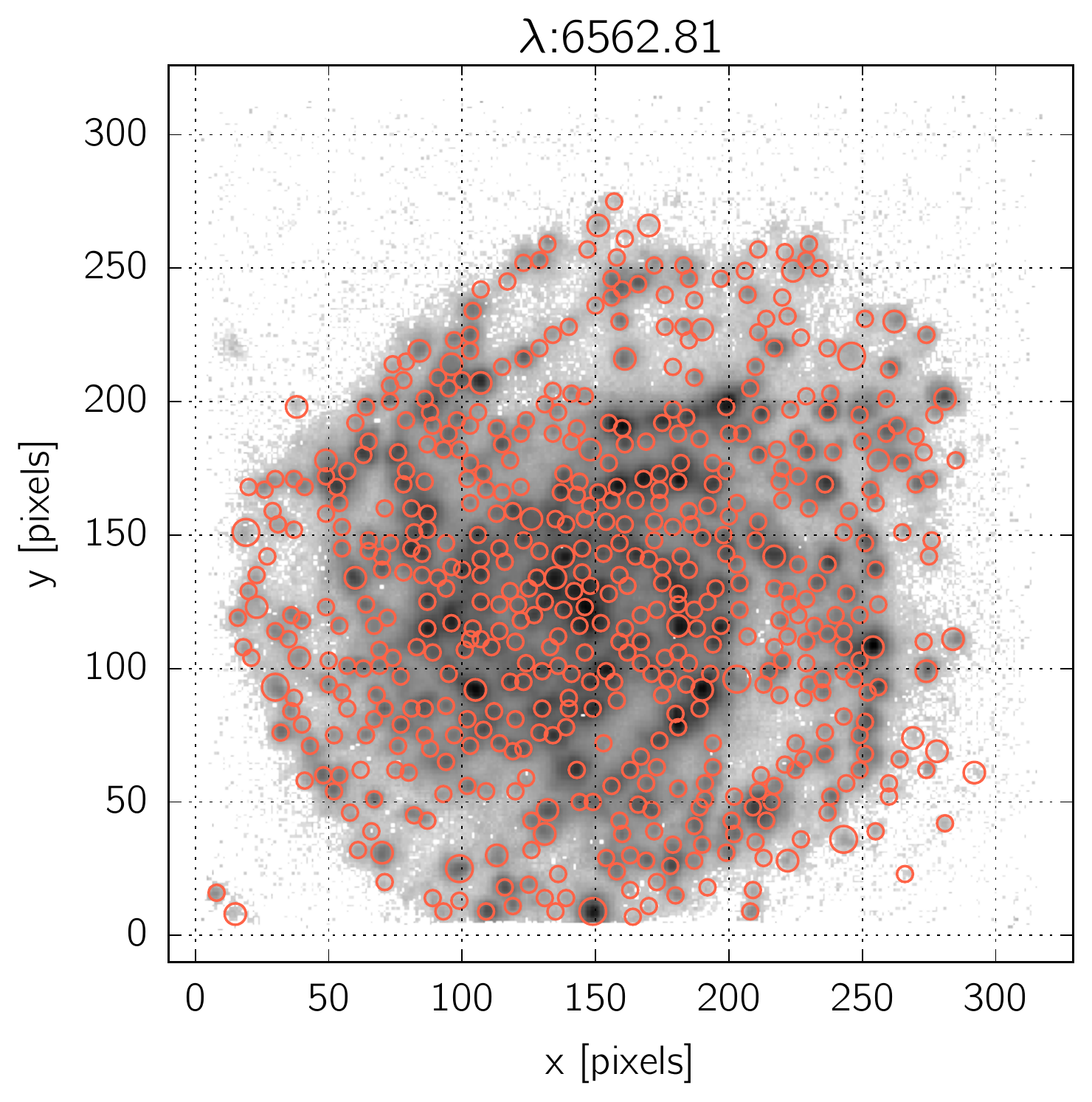}}
\caption{H$\alpha$ flux map of HCG~91c overlaid with the 556 apertures associated with local maxima. The apertures were first identified automatically by \textsc{brutus}, and subsequently inspected and adjusted manually using a dedicated interactive module inside the code.}
\label{fig:apertures}
\end{figure}

Whereas a spaxel-based analysis best exploits the high spatial resolution of MUSE observations, the SNR in the strong emission lines (and in particular H$\beta$) is too little for numerous star-forming regions in the outskirts of the galaxy to perform a reliable analysis. Hence, we supplement the spaxel-based approach with an aperture-based one. Namely, we use \textsc{brutus} to detect H\,\textsc{\smaller II} regions in the data cube automatically, by detecting local maxima in the integrated H$\alpha$ flux map (see Fig.~\ref{fig:flux-maps}). Our approach is somewhat reminiscent of that adopted by \cite{Sanchez2012}, but our codes are entirely different in practice (see Appendix~\ref{app:brutus} for details). For HCG~91c, we defined 556 circular apertures associated with individual local maxima in the H$\alpha$ flux map, all of which are presented in Fig.~\ref{fig:apertures}. 

\subsection{Deriving the oxygen abundance and ionization parameter estimates}

From the set of strong emission line fluxes derived from both the spaxel-based and aperture-based spectral fitting, we derive estimates of the gas-phase oxygen abundance 12+$\log$(O/H) and ionization parameter $\log$(Q). Here, we rely on the \textsc{pyqz} \textsc{python} module v0.8.1, first introduced (as v0.4) in \cite{Dopita2013} with the full propagation of observational errors subsequently implemented in v0.6 \citep[see Appendix B in][]{Vogt2015}. The latest embodiment of the code relies on the photo-ionization models from the \textsc{mappings v} code\footnote{\url{https://miocene.anu.edu.au/Mappings/}} (Sutherland et al., in prep.), and is now hosted on a dedicated Github repository\footnote{\url{http://fpavogt.github.io/pyqz}}. The principle and physical basis of the code however remains unchanged with respect to \cite{Dopita2013}: specific line ratio spaces --in which the photo-ionization model grids are \emph{flat, without wraps} and cleanly allow to disentangle the influence of 12+$\log$(O/H) and $\log$(Q)-- are used to derive estimates of these two parameters associated with a given set of strong line fluxes. For this analysis, we rely on the plane-parallel photo-ionization models with $P_k=5.0$ and $\kappa=\infty$ \citep[i.e. a Maxwell-Boltzmann energy distribution for the electrons, see][]{Nicholls2012,Nicholls2013}. The propagation of observational errors is achieved through the generation of 400 (in this case) random realizations of line fluxes according to the error distribution of the measured lines (assumed to be gaussian), and the subsequent reconstruction of the full probability density function in the 12+$\log$(O/H) \emph{vs.} $\log$(Q) plane from these 400 individual estimates, using Kernel Density Estimate (KDE) techniques. For this work, we adopt the \textsc{mappings} 5.1 models with the local Galactic concordance (LGC) abundance set, in which the \textit{local region} reference abundance corresponds to 12+$\log$(O/H) = 8.76 (Nicholls et al., 2016, MNRAS, submitted).

Several of the \textsc{pyqz} diagnostic grids involve the [O\,\textsc{\smaller II}]$\lambda\lambda$3726,3729 emission lines, which at the redshift of HCG~91c do not fall within the MUSE spectral range. Here, we derive our estimates of 12+$\log$(O/H) and $\log$(Q) from the combination of the following two diagnostic grids that do not employ that line:
\begin{eqnarray}
\log \left( \frac{[\mathrm{N}\,\textsc{\smaller II}]\lambda 6583}{[\mathrm{S}\,\textsc{\smaller II}]\lambda\lambda 6716, 6731}\right) &vs& \log\left( \frac{[\mathrm{O}\,\textsc{\smaller III}]\lambda 5007}{\mathrm{H}\beta} \right),\mathrm{ and} \nonumber \\
\log \left( \frac{[\mathrm{N}\,\textsc{\smaller II}]\lambda 6583}{[\mathrm{S}\,\textsc{\smaller II}]\lambda\lambda 6716, 6731}\right) &vs& \log\left( \frac{[\mathrm{O}\,\textsc{\smaller III}]\lambda 5007}{[\mathrm{S}\,\textsc{\smaller II}]\lambda\lambda 6716,6731} \right).
\end{eqnarray}

\begin{figure}
\centerline{\includegraphics[width=\hsize]{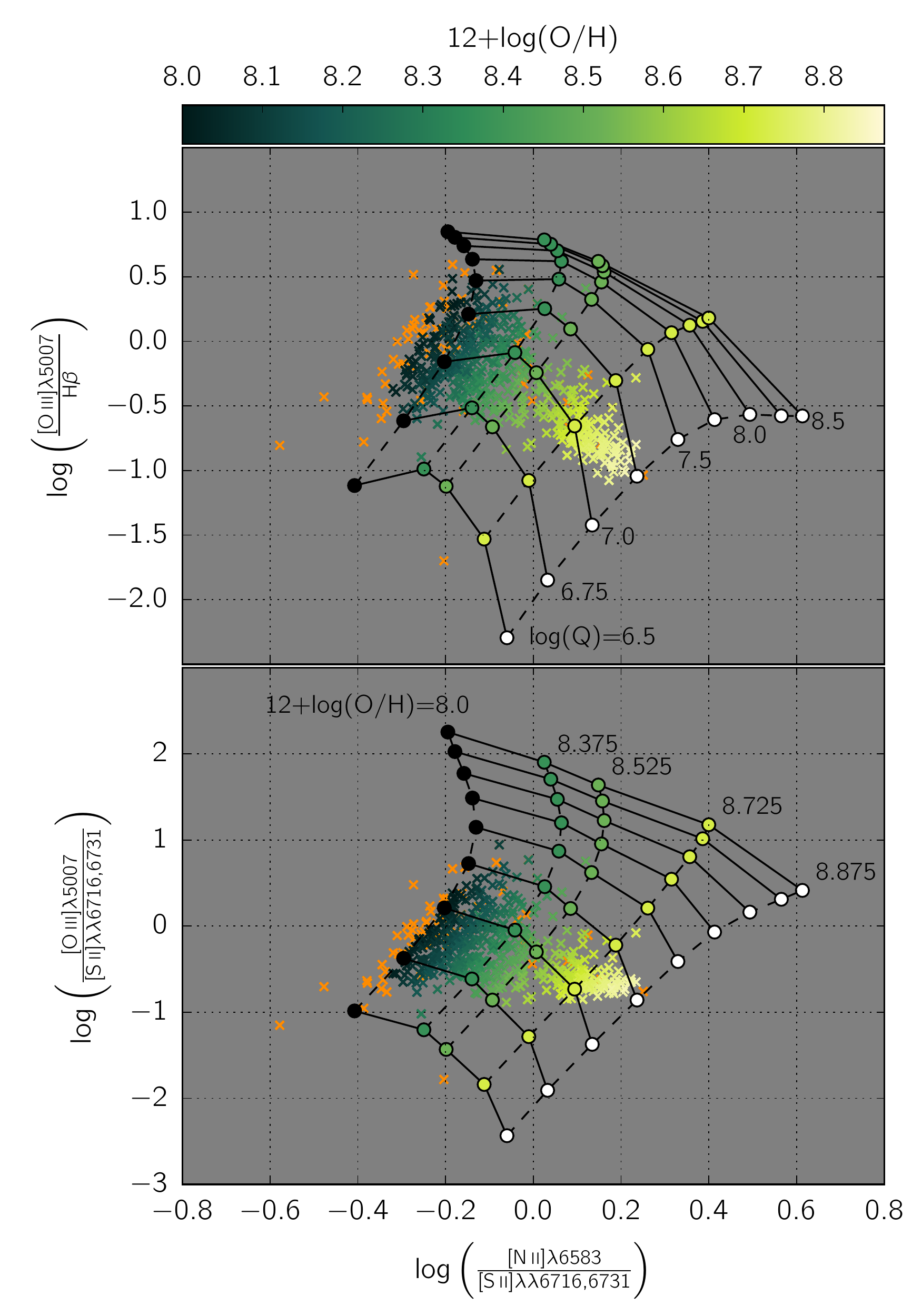}}
\caption{Emission line ratio diagnostic grids from the \textsc{mappings} 5.1 photo-ionization code, for a plane-parallel geometry with $P_k=5.0$ and $\kappa=\infty$. Each circle corresponds to one simulation with a specific abundance and ionization parameter. Small crosses indicate the location of each of the 556 circular aperture, color-coded as a function of the combined (where suitable) abundance value derived from both diagnostics. Orange crosses mark the apertures for which \textsc{pyqz} could not derive a reliable abundance: either from a too large discrepancy between both grids, or because the estimates fall outside both grids.}
\label{fig:pyqz_grids}
\end{figure}

The distribution of each of the 556 apertures in both line ratio planes in shown in Fig.~\ref{fig:pyqz_grids}. Both diagnostics are in excellent agreement throughout most of the abundance range --except for the star-forming regions with 12+$\log$(O/H)$\lesssim 8.1$, which tend to be less enriched according to the first diagnostic grid. This mismatch is most likely impacted by problems within the theoretical models at the low-abundance end, rather than fully by a miscorrection of the extragalactic reddening (which is nonetheless likely to be playing a role as well). Indeed, a reddening correction issue would have likely affected all apertures, which is not seen here.

A detailed investigation of this mismatch --complicated by the lack of the [O\,\textsc{\smaller II}]$\lambda\lambda$3726,3729 in the MUSE spectral range-- is outside the scope of this article. It may for example be that our model with $P_k=5.0$ may not be suitable for these star-forming regions, and/or this mismatch may simply reflect genuine limitations of the \textsc{mapping v} models at the low abundance end. In this analysis, we choose to also use the apertures for which only one diagnostic grid returns a suitable estimate, noting and stressing that this choice does not affect our conclusions. In particular, we note that the trends discussed in the next Section are present when considering either diagnostic grids separately, or together.

\section{Results}\label{sec:results}

\begin{figure}
\centerline{\includegraphics[width=\hsize]{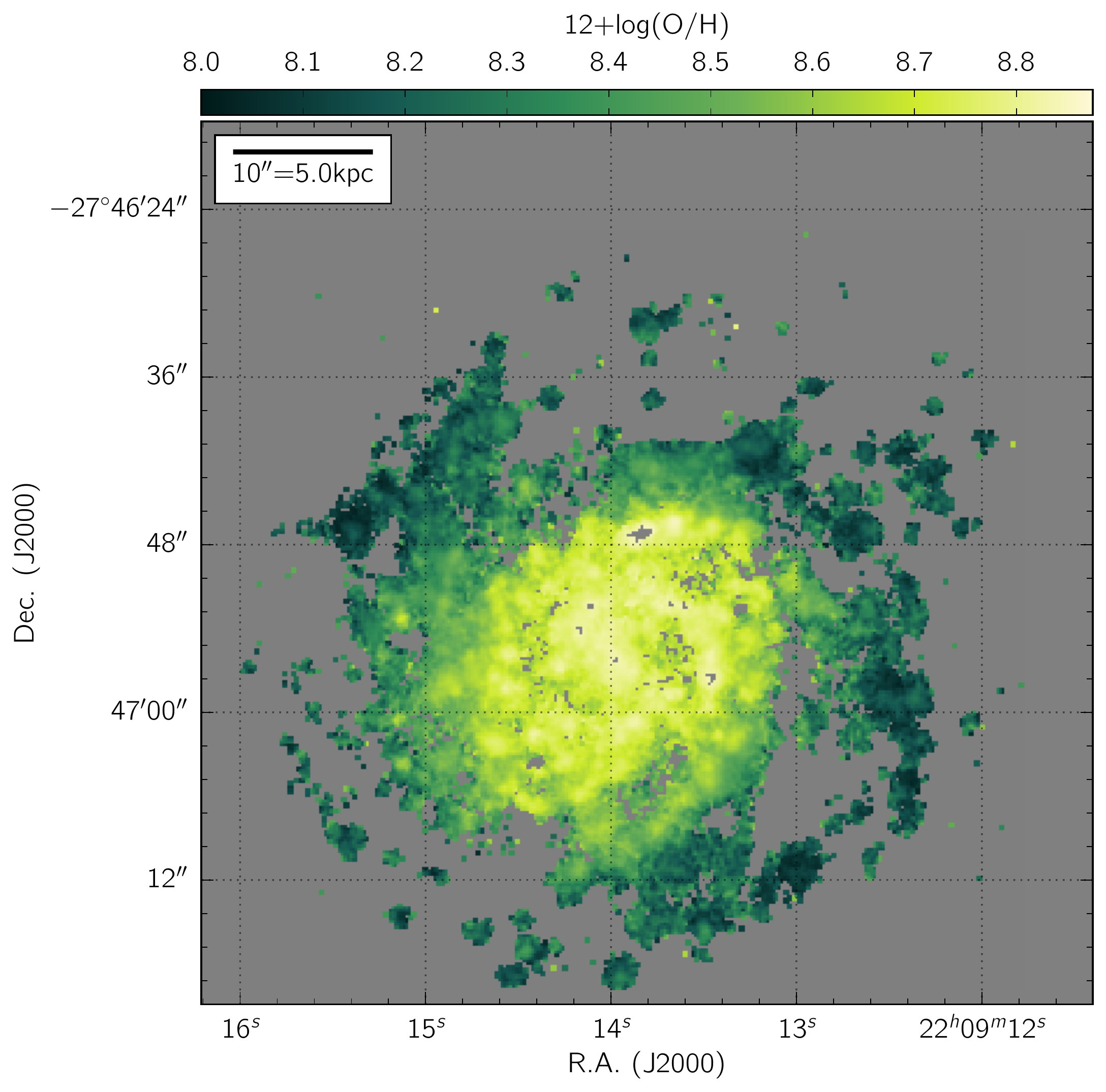}}
\centerline{\includegraphics[width=\hsize]{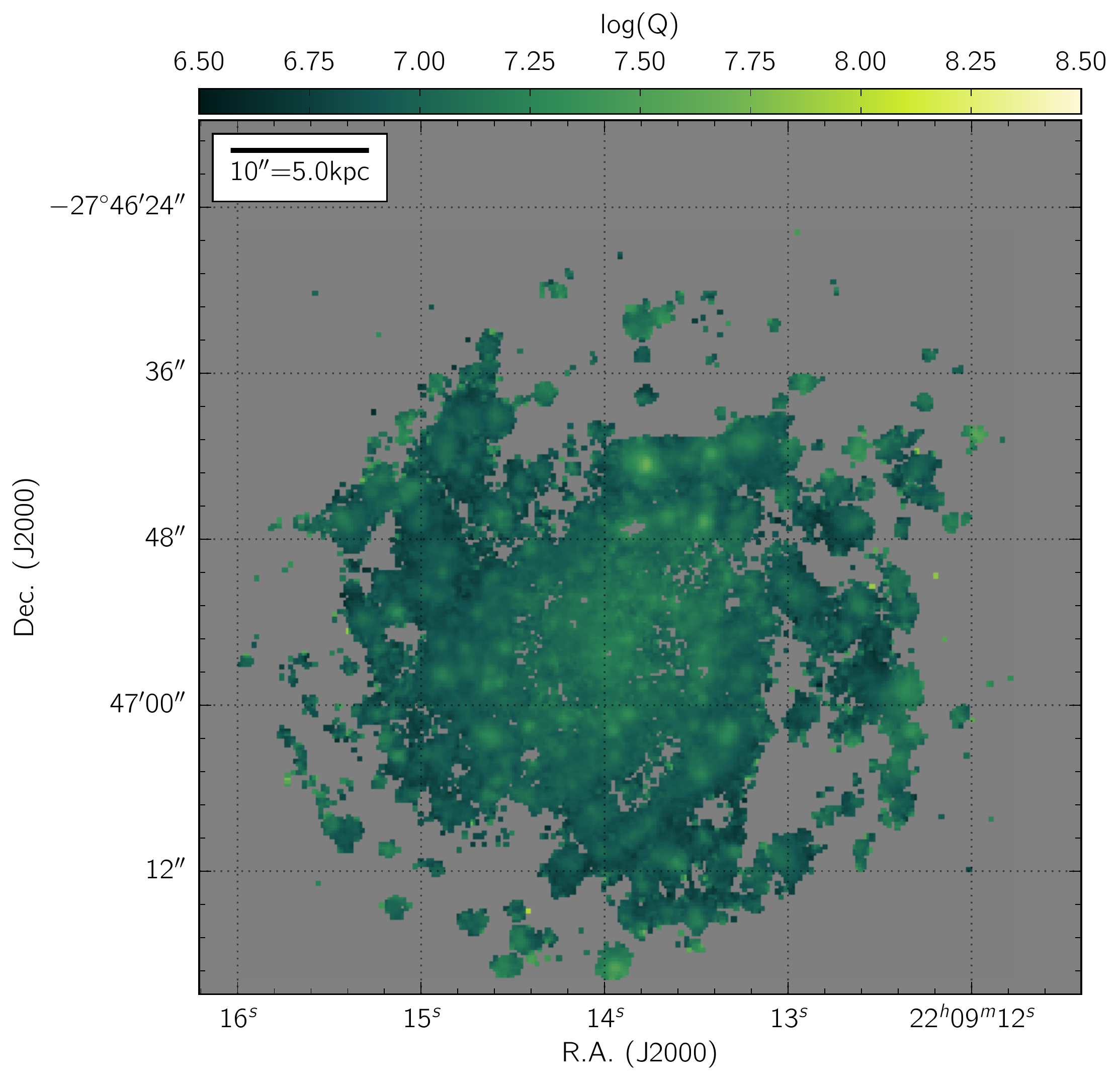}}
\caption{Spaxel-based map of the oxygen abundance (top) and ionization parameter (bottom) in HCG~91c, constructed using the \textsc{brutus} and \textsc{pyqz} codes. The color bars span the full range of values covered by the \textsc{mappings v} simulations, highlighting the large range of oxygen abundances and narrower range of ionization parameters throughout the galaxy.}
 \label{fig:Z-maps}
 \end{figure}

\begin{figure}
\centerline{\includegraphics[width=\hsize]{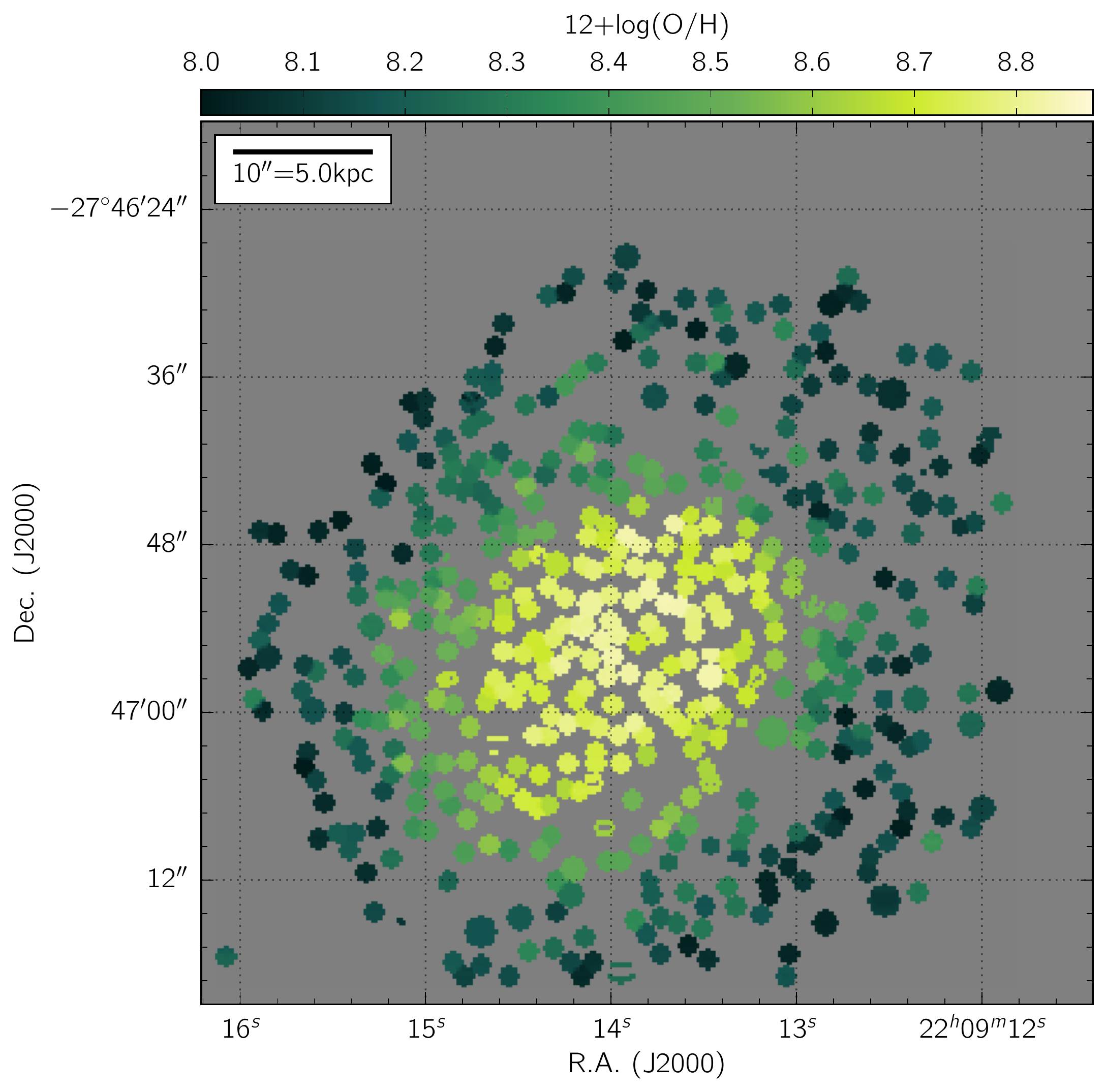}}
\centerline{\includegraphics[width=\hsize]{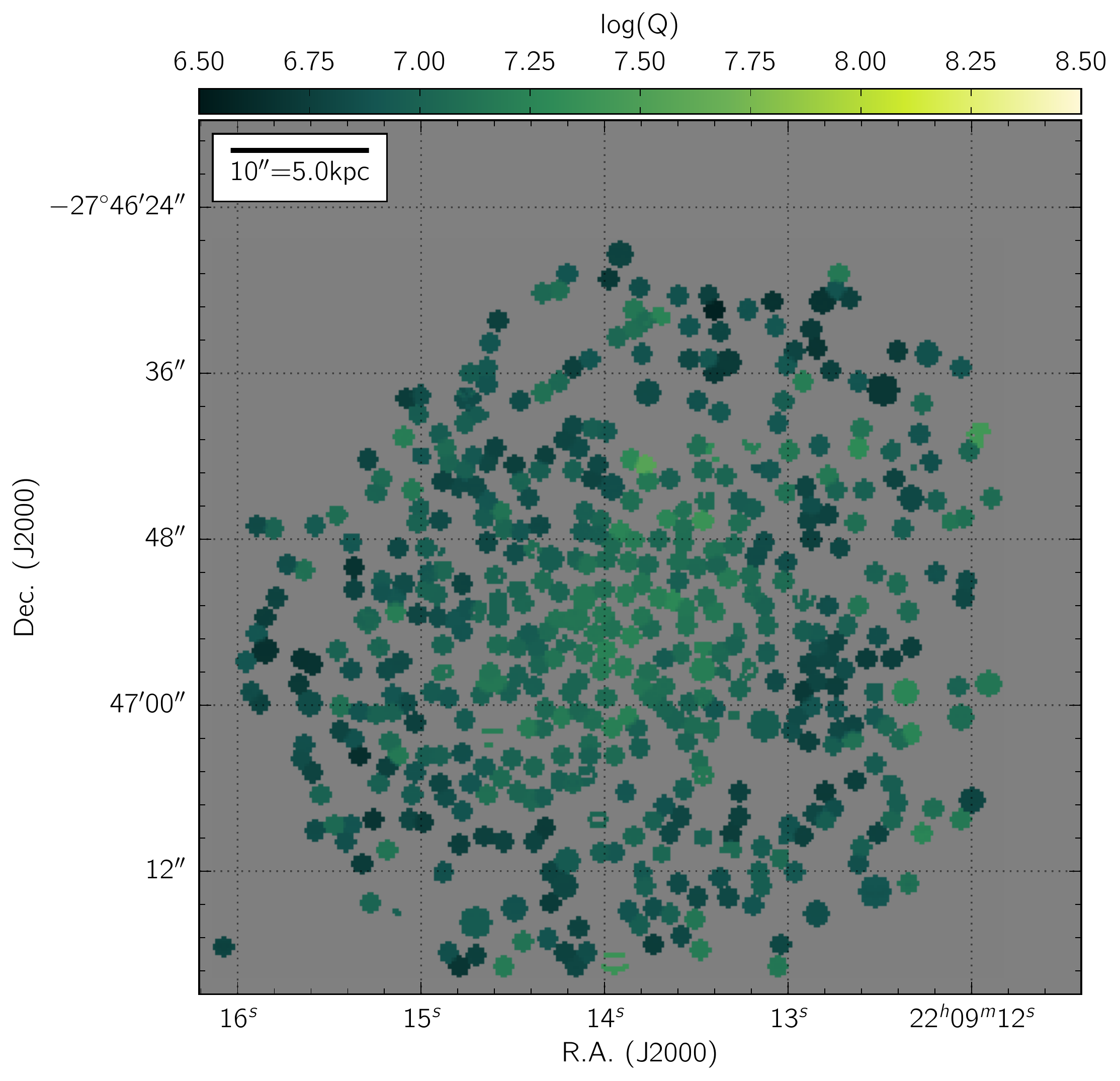}}
\caption{Same as Fig.~\ref{fig:Z-maps}, but for the 556 apertures associated with local maxima in the H$\alpha$ flux map of the galaxy, identified with \textsc{brutus}.}
 \label{fig:Q-maps}
 \end{figure}

The maps of 12+$\log$(O/H) and $\log$(Q) for both the spaxel-based case and the aperture-based case are presented in Figs.~\ref{fig:Z-maps} and \ref{fig:Q-maps}. The spatial resolution of these maps (and the gain provided by MUSE) can be compared with those obtained with the WiFeS integral field spectrograph described in \cite{Vogt2015}: the general features identified with WiFeS remain (in particular the sharp decrease in the oxygen abundance at $\sim$6\,kpc North from the galaxy center), but the significant improvement in the spatial resolution (both in terms of seeing and spaxel size) also allows to better resolve structures within the disk, as well as to detect and characterize H\,\textsc{\smaller II} regions at larger radii than with WiFeS. 

Local enhancements of the ionization parameter are present throughout the disc, and are associated with individual star-forming complexes. The largest values of $\log$(Q) are found in the outskirts of the galaxy, beyond the effective radius ($R_{e}$=5.1\,kpc). In terms of the oxygen abundance, the picture revealed by MUSE is clearly more complex than that discussed by \cite{Vogt2015} from the WiFeS observations of the system. 

\begin{figure*}
\centerline{\includegraphics[scale=0.5]{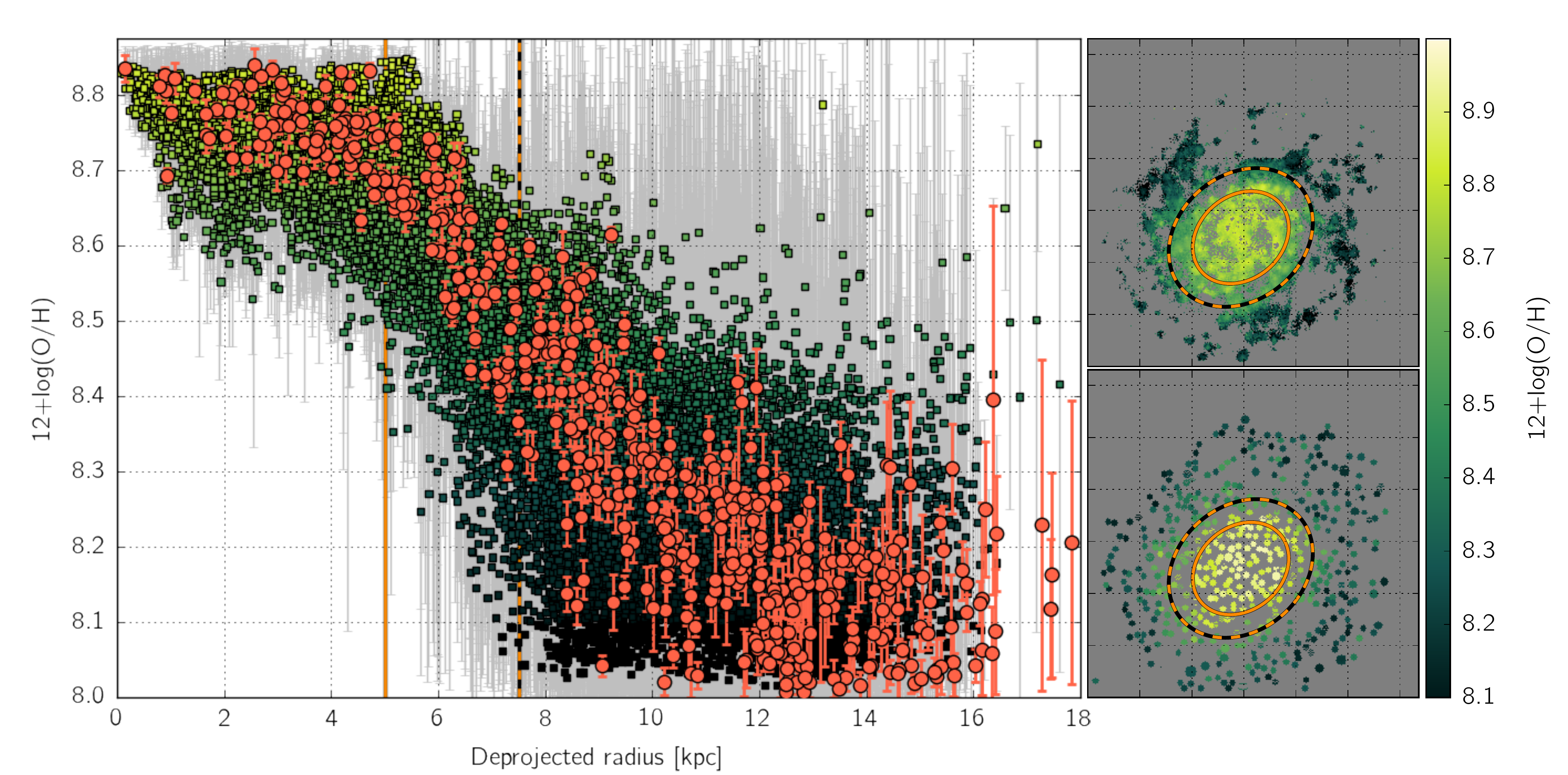}}
\centerline{\includegraphics[scale=0.5]{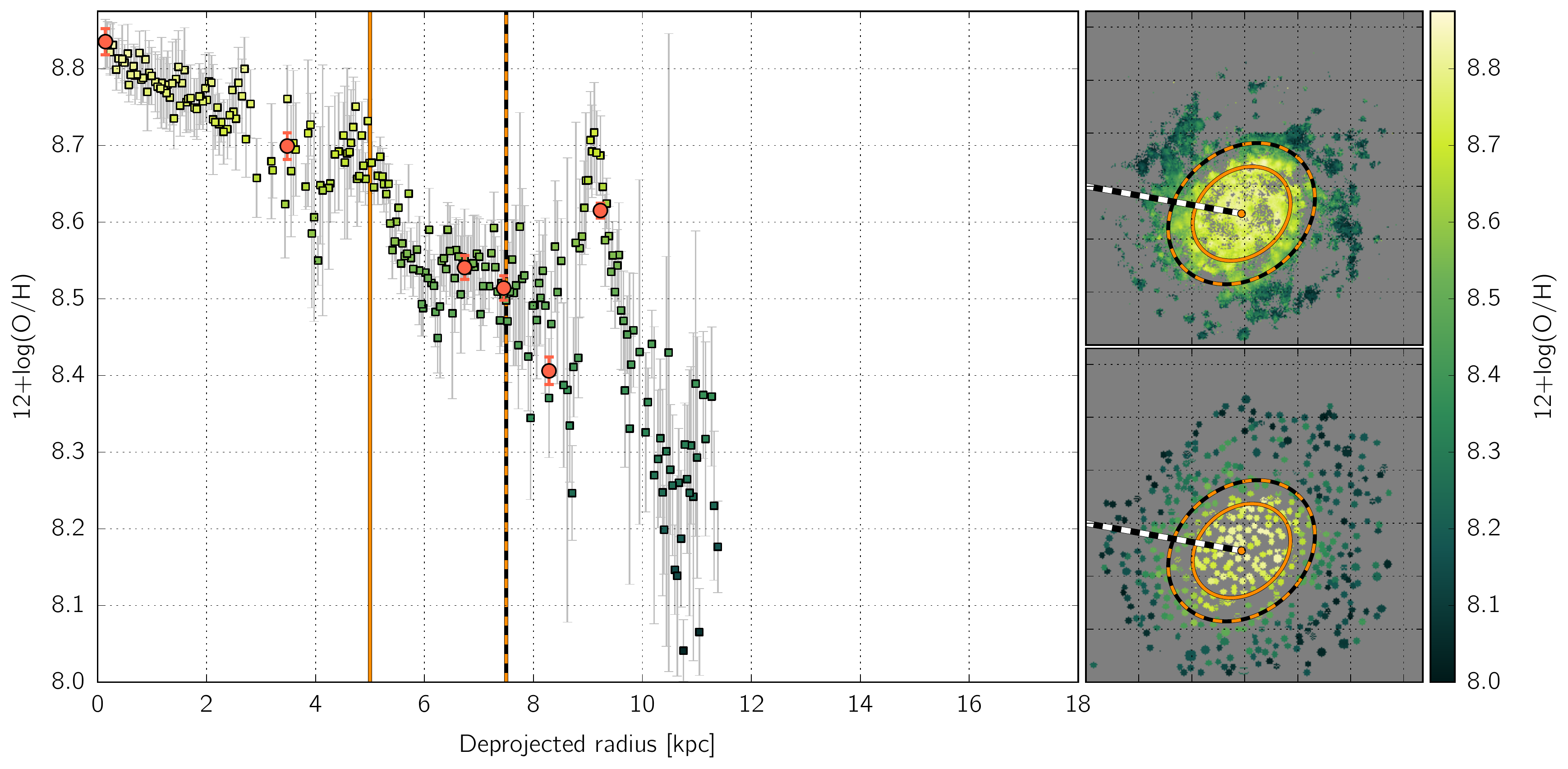}}
\caption{Global oxygen abundance gradient (top panel) and for regions within 0.3\,arcsec of the specific azimuth 79.5$^{\circ}$ East-of-North (i.e. the area between the white \& black dashed line). Individual spaxels are shown as colored squares with the associated 1-$\sigma$ errors in grey. Measurements for apertures (with associated 1-$\sigma$ errors) are shown in red. Whereas globally, the gradient in HCG~91c can be described as linear with a flattening inwards of 4\,kpc and (possibly) outwards of 11\,kpc, the trends are highly non-linear for individual azimuths. Rapid and localized variations, detected both in individual spaxels and integrated apertures alike are present at all radii. Ellipses marking the deprojected effective radius R$_{e}$ (5.1\,kpc, orange) and 1.5\,R$_{e}$ (dashed-orange) radius from the galaxy center are shown in all oxygen abundance maps (both spaxel-based and aperture based).}
 \label{fig:gradient}
 \end{figure*}

\begin{figure*}
\centerline{\includegraphics[scale=0.5]{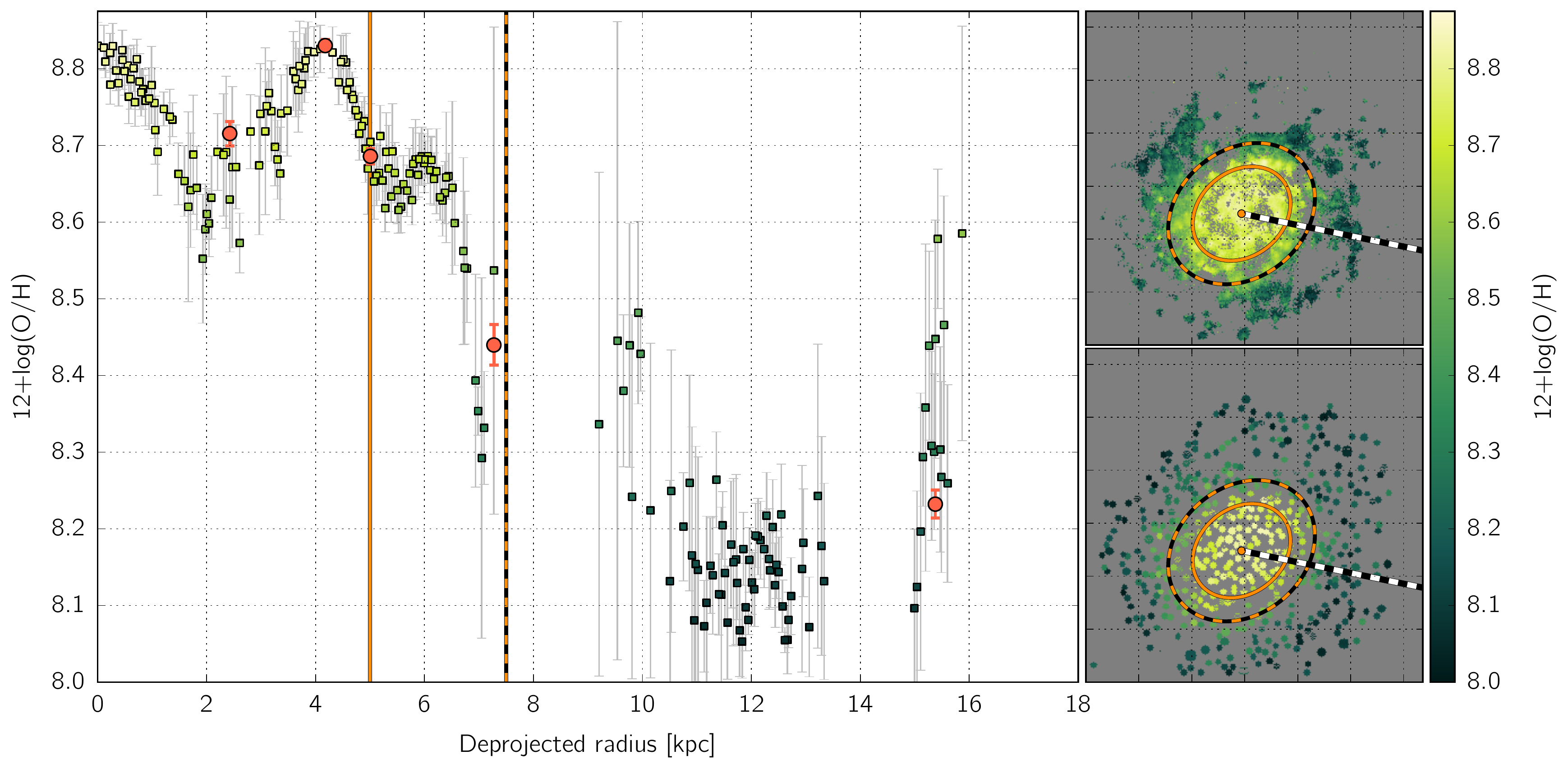}}
\centerline{\includegraphics[scale=0.5]{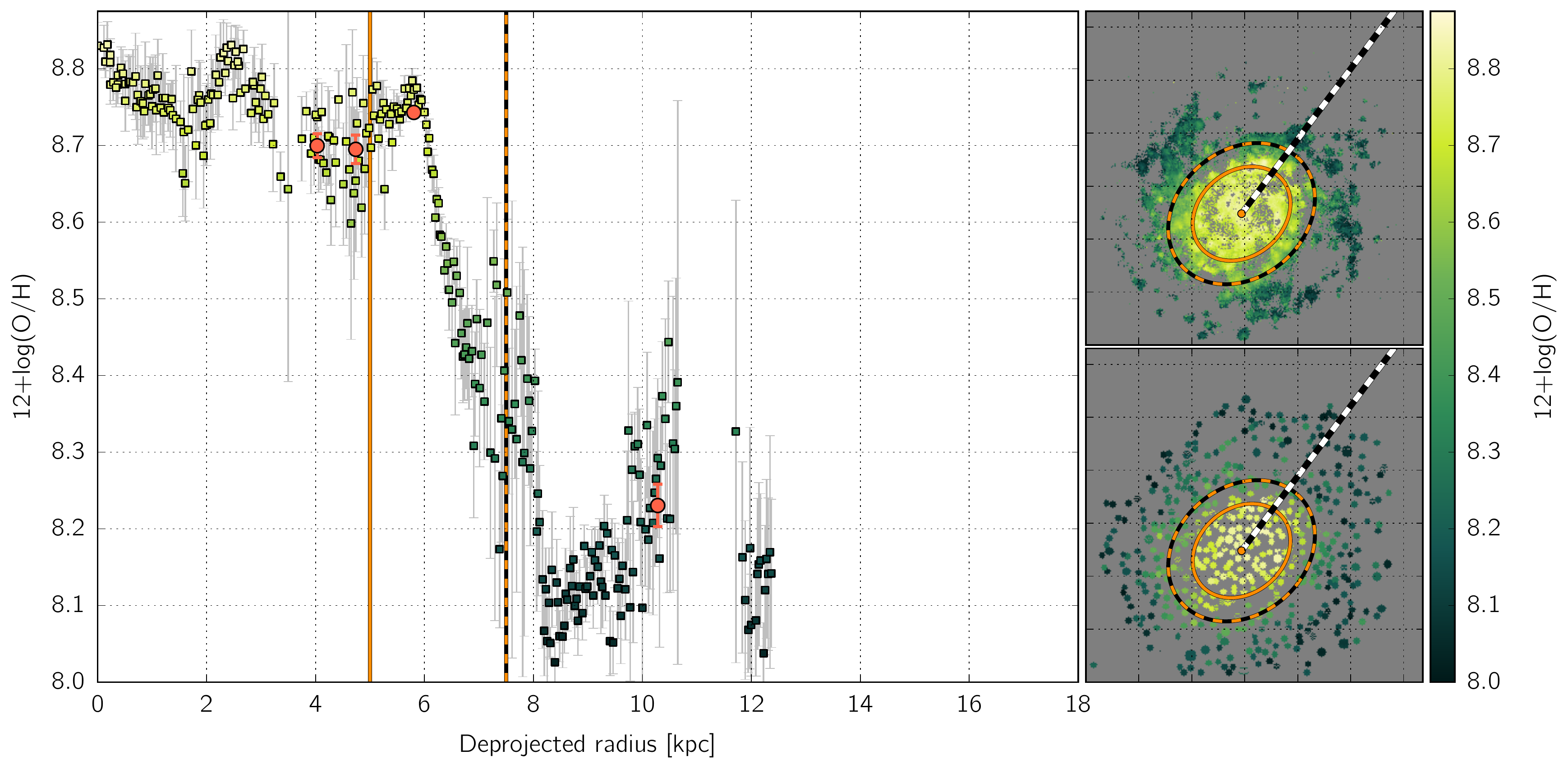}}
\caption{Same as Fig.~\ref{fig:gradient} but for azimuths  258.0$^{\circ}$ (top) and 322.5$^{\circ}$ (bottom) East-of-North, respectively. For completeness, the individual oxygen abundance gradients of HCG~91c extracted for all azimuths in steps of 0.5$^{\circ}$ have been stacked into a movie available as supplementary material. \textit{Note to arXiv readers: until publication, the movie will be available at \texttt{http://www.sc.eso.org/$\sim$fvogt/supp\_mat/HCG91c/O\_gradient.mp4}}}
 \label{fig:azimuth}
 \end{figure*}

We present in Fig.~\ref{fig:gradient} and \ref{fig:azimuth} the oxygen abundance gradient of HCG~91c observed by MUSE, both global and along specific azimuths. When considering all spaxels and/or apertures at once, the oxygen abundance in HCG~91c displays a linear decline of $-0.082\pm0.001$\,dex\,kpc$^{-1}\equiv -0.418\pm0.005$\,dex\,R$_{e}^{-1}$ from 1\,R$_{e}$ (5.1\,kpc) to 2\,R$_{e}$ (10.2\,kpc), with a flattening inwards of $\sim$0.8\,R$_{e}$ (4\,kpc), and (possibly) outwards of 2.1\,R$_{e}$ (11\,kpc): a trend already already detected with WiFeS \citep{Vogt2015}. Modulo a reduced scatter, both the spaxel-based and aperture-based gradients display similar trends, indicative that (as one would expect) the distance to HCG~91c is large enough not to affect a spaxel-based approach through the resolution of the temperature structures of H\,\textsc{\smaller II} regions. The consistency between the spaxel-based and aperture-based approach also indicates that the presence of Diffuse Ionized Gas (DIG) in HCG~91c is not affecting our ability to derive reliable estimates of the 12+$\log$(O/H) and $\log$(Q). We note that large areas void of clearly-identified star-forming regions (e.g. the inter-arm region to the South-West of the galaxy center) -and thus possibly dominated by DIG emission- have too little S/N to be processed reliably by \textsc{pyqz}.

\begin{figure}
\centerline{\includegraphics[scale=0.5]{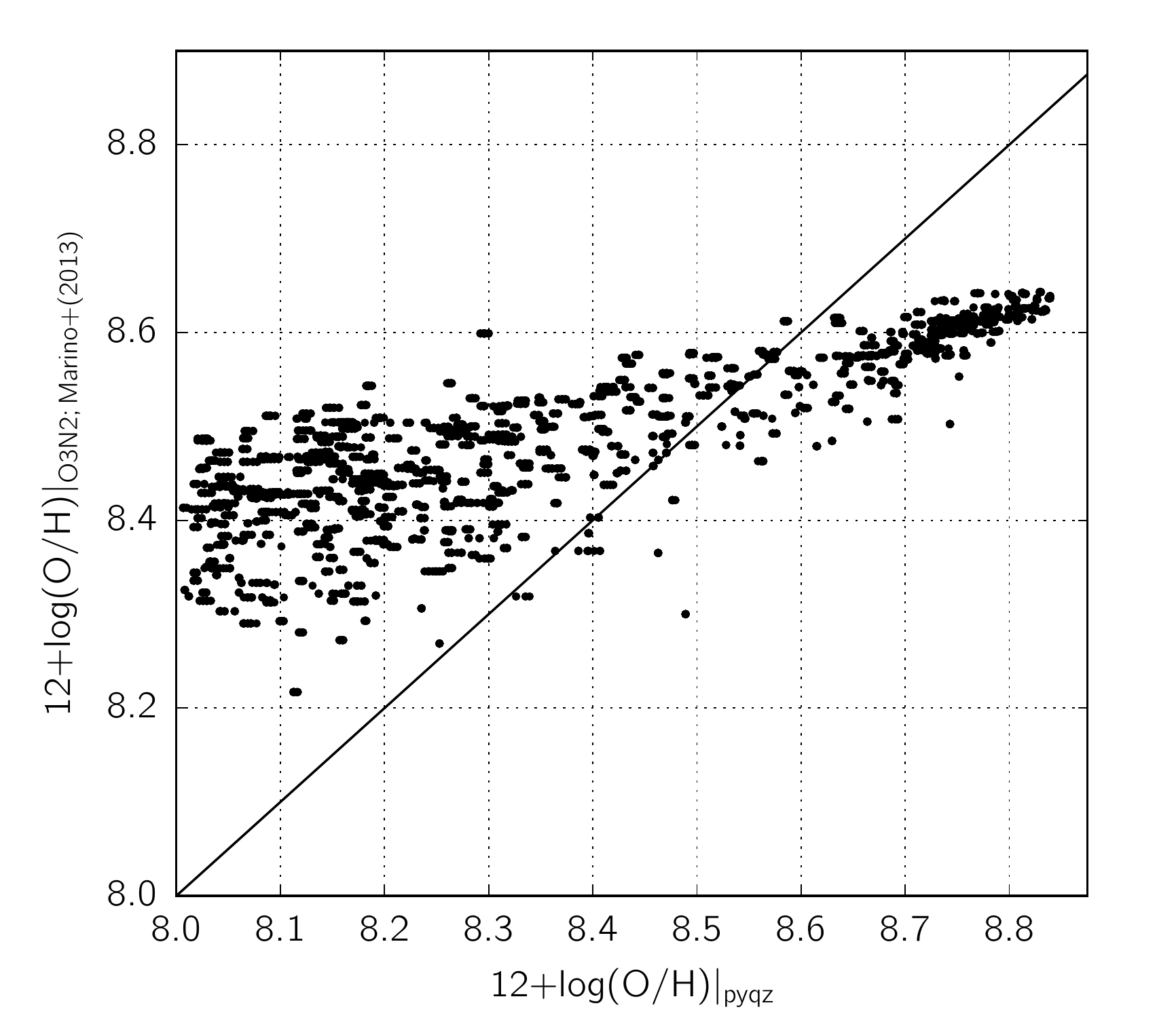}}
\caption{Comparison of the oxygen abundance values derived via \textsc{pyqz} v0.8.1 (i.e. using the \textsc{mappings v} photo-ionisation models; this work) and the O3N2 calibration of \cite{Marino2013}, as employed by \cite{Sanchez-Menguiano2016} for all the apertures spectra in HCG~91c. \textsc{pyqz} estimates span a wider range of oxygen abundances, so that \textsc{pyqz}-derived gradient slopes are steeper than those derived by \cite{Sanchez-Menguiano2016}. Despite a scatter of $\sim$0.1\,dex at the lower abundance end, the correspondence between the two techniques remains nonetheless linear over the span of oxygen abundances in HCG 91c.}
 \label{fig:O3N2}
 \end{figure}

A flattening of the oxygen abundance gradient beyond 2\,R$_{e}$ has been identified as a generic feature in CALIFA galaxies \citep{Sanchez2014,Sanchez-Menguiano2016,Zinchenko2016}, with the flattening of the gradient inside $\sim$0.5\,R$_{e}$ identified primarily in galaxies with stellar masses $\log(M_{\star}/M_{\odot})\geq10.5$. The gradient slope derived by \textsc{pyqz} for HCG~91c is somewhat steeper than the universal gradient measured with CALIFA galaxies \citep[$-0.075$\,dex\,$R_{e}^{-1}$ with a scatter of 0.016\,dex\,$R_{e}^{-1}$, see][]{Sanchez-Menguiano2016}. This mismatch is due to the different techniques used to derive the oxygen abundance values (as illustrated in Fig.~\ref{fig:O3N2}): compared to the O3N2 calibrations of \cite{Marino2013}, \textsc{pyqz} and the underlying \textsc{mappings v} photo-ionization models return a wider range of abundances in HCG~91c, leading to a steeper gradient. Taking this scaling difference into account, the oxygen abundance gradient in HCG~91c (between 1\,R$_{e}$ and 2\,R$_{e}$) is consistent with the universal gradient reported by \cite{Sanchez-Menguiano2016}.
 
Our ability to spatially resolve sub-kpc scales for all azimuths in HCG~91c also reveals additional features of the oxygen abundance distribution throughout the galaxy, beyond the existence of a global gradient. Two distinct types of behaviors are present: first, on sub-kpc-scales, and second on kpc-scales. We describe them separately in the next Sections.

\subsection{The sub-kpc-scale variations of 12+log(O/H)} 

The global gradient visible in Fig.~\ref{fig:gradient} (top) contains a vertical scatter of $\sim$0.2\,dex in the oxygen abundance present at all radii in HCG~91c. When considering only the aperture-based measurements, the vertical scatter remains of the order of $\sim$0.15\,dex despite intrinsically smaller measurement errors. It becomes evident that this vertical scatter is real (rather than related to measurement errors) when inspecting the oxygen abundance gradient as a function of the azimuth, as illustrated by the lower panel of Fig.~\ref{fig:gradient} and those in Fig.~\ref{fig:azimuth}. Restricting the azimuthal range of the gradient diagram reveals the presence of spatially localized and coherent variations in the measured oxygen abundances. The variations are of the order of 0.1-0.2\,dex over distances $\leq$1\,kpc, significant at more than 5-$\sigma$ for most cases, and detected both in individual spaxels and integrated apertures (both consistent with one another). These rapid variations can be visually identified in the maps of the oxygen abundance (see Fig.~\ref{fig:Z-maps}), the clearest example of which is located 15\,arcsec East of the galaxy center.  

\subsection{The kpc-scale variations of 12+log(O/H)} 

\begin{figure*}
\centerline{\includegraphics[width=\textwidth]{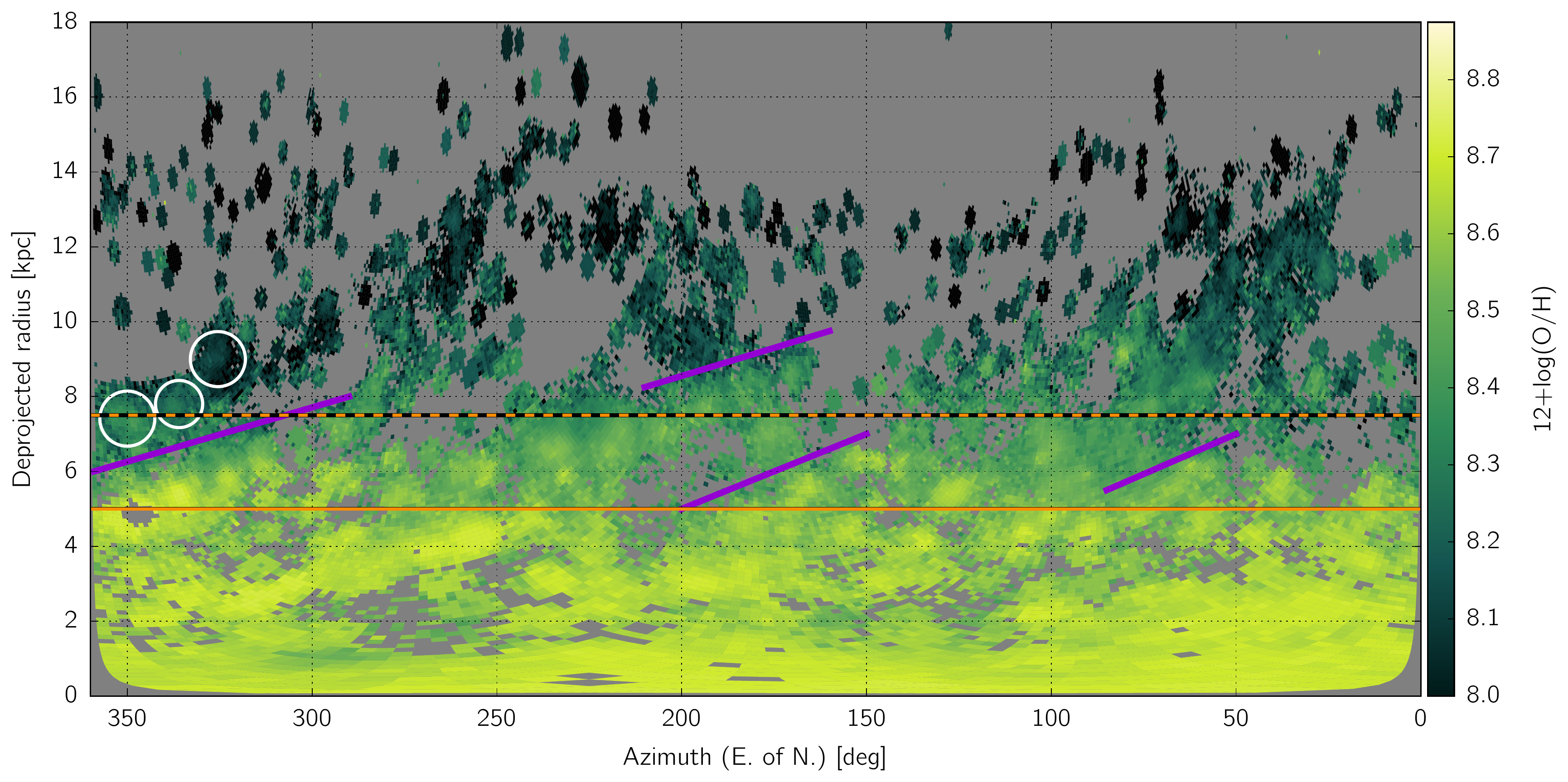}}
\caption{The \emph{de-wrapped} abundance distribution of HCG~91c. Each spaxel in the datacube was reprojected in the azimuth-distance plane according to its location with respect to the galaxy center. For spaxels associated to a given aperture but without a reliable \emph{individual} measure of 12+$\log$(O/H), the aperture-derived value of 12+$\log$(O/H) is shown instead. Spiral arms can be identified and tracked from 2\,kpc outwards in this de-projected space. The iso-distance ellipses shown in the different panels of Fig.~\ref{fig:gradient} and Fig.~\ref{fig:azimuth} become horizontal line in this projection. Inclined purple lines mark the boundaries of spiral structures displaying a rapid variation of the oxygen abundance. The star-forming complexes found by \cite{Vogt2015} to have discrepant oxygen abundances with respect to their immediate surroundings are marked with white circles.}
\label{fig:dewrapped}
\end{figure*}

An alternative approach to visualizing the gaseous oxygen abundance distribution in HCG~91c is presented in Fig.~\ref{fig:dewrapped}, where we \textit{de-wrapped} the oxygen abundance gradient on the \emph{azimuth-distance} plane. In this projection, all points at a given height are located at the same distance from the galaxy center. This projection facilitates the identification and inspection of spiral structures beyond the effective radius, at the cost of lesser clarity closer from the galaxy center (stretched horizontally across the diagram).
 
\begin{figure}
\centerline{\includegraphics[width=\hsize]{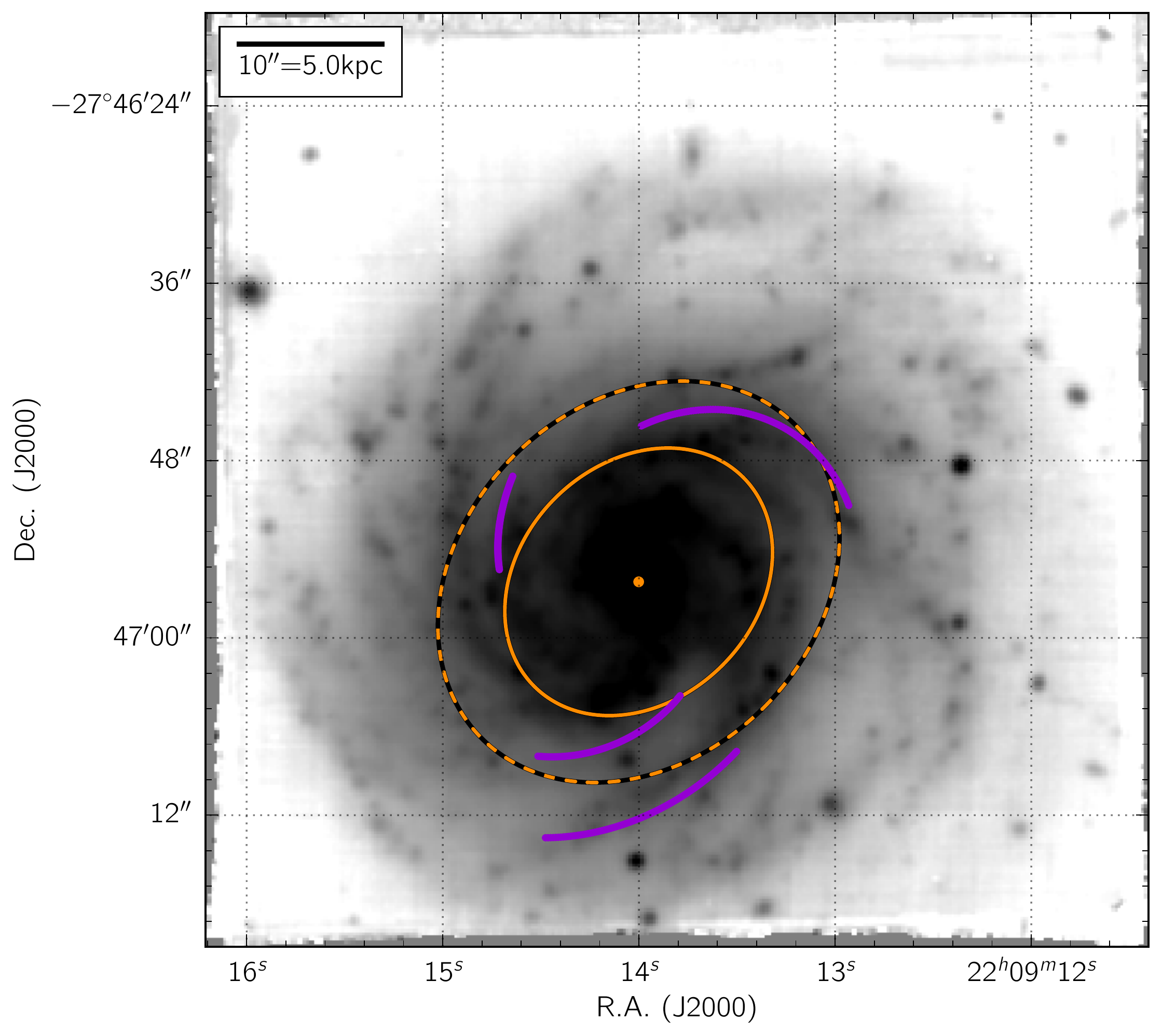}}
\caption{White-light image of HCG~91c, reconstructed by collapsing the entire MUSE datacube. The galaxy center is marked with an orange dot. The projected effective radius $R_{e}$ and $1.5\cdot R_{e}$ ellipses are shown in orange and dashed-orange, respectively. The location of the coherent, kpc-scale variations of the oxygen abundance traced in Fig.~\ref{fig:dewrapped} are shown with purple curves. These effectively trace the edge of several of the spiral structures of HCG~91c.}
\label{fig:spiral_arms}
 \end{figure}

Large kpc-scale coherent trends in the oxygen abundance distribution pattern are present throughout HCG~91c, in addition to the spatially localized variations described previously. These features, the most prominent of which are traced by purple lines in Fig.~\ref{fig:dewrapped}, are located at the edges of the spiral arms of the galaxy, as illustrated in Fig.~\ref{fig:spiral_arms}. In other words, the variations in the gaseous oxygen abundance are sharper and more abrupt when moving across the spiral structure, and more gradual when moving along the spiral arms. The rapid variation of the oxygen abundance measured with WiFeS to the North of the galaxy center \citep{Vogt2015} corresponds to the (then) unresolved \emph{crossing} of a spiral structure at that location, which the MUSE datacube reveals is not an isolated case, but rather the sharpest example of a behavior present at all azimuths.

\section{Discussion}\label{sec:discussion}

The detection of sub-kpc scale, spatially localized variations of the oxygen abundance in HCG~91c --a spiral galaxy-- is not excessively suprising. In near-by grand design spirals for example, \cite{Bresolin2009} report that the intrinsic scatter of the oxygen abundance is of the order of 0.2\,dex throughout the disk. \cite{Croxall2016} measure intrinsic dispersion in 12+$\log$(O/H) of 0.074\,dex in NGC~5457 \citep[M101, see also][]{Kennicutt1996}. In the same galaxy, \cite{Li2013} report two cases of locally lower and higher oxygen abundances for two H\,\textsc{\smaller II} regions compared to their immediate surroundings. \cite{Rosolowsky2008} detect a scatter of 0.21\,dex in M33 \citep[but see also][that from a sample of 25 H\,\textsc{\smaller II} regions only finds a scatter of 0.06\,dex]{Bresolin2011}. \cite{Sanders2012} find localized variations of the oxygen abundance associated with individual star-forming regions of the order of 0.3\,dex in M31, and \cite{Berg2015} report a scatter of 0.165\,dex in NGC~0628 \citep[see also][]{Rosales-Ortega2011}. Although caution must certainly be used when comparing these different studies with one another, the presence of localized variations of the oxygen-abundance of star-forming regions with respect to their immediate surroundings within spiral galaxies remains unequivocal. From this perspective, the case of HCG~91c clearly illustrates the importance of characterizing the gaseous oxygen abundance throughout the entire optical discs of galaxies \emph{while} distinguishing individual star-forming complexes within, in order to capture both pc-scale and kpc-scale trends. For long slit studies of near-by systems, the case of HCG~91c reinforces the importance of selection bias in deriving local abundance scatter and global trends from a handful of locations within a galaxy's disc. Ideally, such studies should characterize different galactic-centric radii (including both in and between spiral arms) and azimuths using not one but several H\,{\smaller II} regions in the immediate vicinity of one another for any given location.

The full mapping of the oxygen throughout HCG~91c also reveals large scale azimuthal variations unambiguously associated with galactic structures (i.e. the spiral arms). Observationally, similar behaviors were already reported for different systems, with varying degrees of certainty. For example, \cite{Sanchez2015} detected (with MUSE) possible azimuthal variations in the oxygen abundance gradient of NGC~6754 associated with the spiral pattern of this galaxy \citep[see also the recent re-analysis of this target by][]{Sanchez-Menguiano2016a}. In NCG~5457, \cite{Croxall2016} identify a mild correlation of the oxygen abundance distribution with the spiral arms, from a sample of $\sim$100 H\,\textsc{\smaller II} regions observed with the Large Binocular Telescope. Such a correlation suggests a picture where ISM enrichment occurs preferentially along the spiral structures, rather than across them. HCG~91c lies within a compact group, but its very regular HI envelope is a clear indication that the galaxy has not yet interacted strongly with the group environment \citep{Vogt2015}. It thus appears more likely that the larger scale abundance variations tracing the spiral pattern in HCG~91c are an intrinsic behaviour of the system, rather than a consequence of the environment. In particular, a (set of) self-driven mechanism(s) appear more plausible than the different environment-driven hypothesis discussed in \cite{Vogt2015}.

Still, the question remains: to what extent are gas phase abundance variations in HCG~91c (local or global) affected by dynamics, and to what extent do they reflect an enhanced star formation efficiency within the spiral structure (that keeps the metals in place)? Theoretically, \cite{Grand2016} presents evidence for the ability of spiral patterns to transfer comparatively more metal-rich stars from the inner regions of the galaxy to the outer parts along the spiral arms \citep[see also][]{Grand2012}. As these enriched stars end their lives in comparatively less enriched regions of the galaxy, they could contribute to a local enhancement of the ISM at larger radii. Alternatively, spiral arms in HCG~91c might have merely been acting as gravitational sink, effectively trapping heavy elements while favouring star-formation activity, thus leading to the present distribution of the oxygen abundance in HCG~91c. Differentiating between these scenarios (and others) would require a better understanding of the nature of the spiral structure in HCG~91c \citep[including its stability and coherence over time, see e.g.][]{Dobbs2014,Baba2015}. From that perspective, a combined \emph{gas+stars} analysis of the MUSE observations of HCG~91c bridging gaseous and stellar abundance (and kinematics) appears very indicated, but outside the scope of this article.

\section{Conclusions}\label{sec:conclusion}

Here we have presented MUSE observations of HCG~91c, as a follow-up of observations acquired with the WiFeS integral field spectrograph \citep{Vogt2015}. Using the new \textsc{brutus} tool designed to process the data cubes of integral field spectrographs, we have have measured the oxygen abundance and ionization parameter associated with star-forming regions throughout the disc of HCG~91c out to $\sim$2\,R${e}$, using both a spaxel-based and aperture-based approach. We confirm the presence of rapid abundance variations in the galaxy initially detected with WiFeS. These variations can be separated in two distinct types. First, sub-kpc-scale variations associated with individual star-forming regions, and second, kpc-scale variations correlated with the spiral structure of the galaxy (specifically with the boundaries of spiral arms). The kpc-scale variations thus provide observational evidence that ISM enrichment is preferentially occurring along the spiral structure of HCG~91c, and less easily across it. As per the sub-kpc-scale variations of the oxygen abundance in HCG~91c, they are reminiscent of the behavior of star-forming regions observed in near-by grand design spirals.

The MUSE observations of HCG~91c confirm the unique ability of this instrument to spatially resolve oxygen abundance gradients, characterize intrinsic scatter and map non-radial variations of the abundance of H\,\textsc{\smaller II} regions in galaxies. The instrument's unique combination of a large FoV and small spaxel size \citep[soon to be fully exploited with the GALACSI ground-layer adaptive optics system and the Four-Laser Guide Star Facility, see][]{Stuik2006,Calia2014} is effectively pushing the distance (out to $\sim$100\,Mpc) at which abundance maps can both cover the optical discs of galaxies out to $\geq2\text{R}_{e}$ while resolving sub-kpc structures. Such targets form the ideal link between projects targeting extremely near-by systems down to pc-scales \citep[e.g.][]{Kreckel2016} to the high redshift Universe: provided observations are performed under excellent seeing conditions, and with exposure times suitable to detect all strong emission lines with sufficient S/N to characterize H\,\textsc{\smaller II} regions up to 2\,R$_{e}$ and beyond.
  
\begin{acknowledgements}

We thank the anonymous referee for a prompt and constructive report.
This research has made use of \textsc{brutus}, a Python module to process data cubes
from integral field spectrographs hosted at \url{http://fpavogt.github.io/brutus/}.
\textsc{brutus} relies on \textsc{statsmodel} \citep{Seabold2010},
\textsc{ppxf} \citep{Cappellari2004}, \textsc{fit\_kinematic\_pa} as described in
Appendix C of \cite{Krajnovic2006}), \textsc{matplotlib} \citep{Hunter2007},
\textsc{astropy}, a community-developed core Python package for Astronomy
\citep{AstropyCollaboration2013}, \textsc{photutils}, an affiliated package of
\textsc{astropy} for photometry, \textsc{aplpy}, an open-source plotting package for
Python hosted at \url{http://aplpy.github.com}, \textsc{montage}, funded by the
National Science Foundation under Grant Number ACI-1440620 and previously funded by
the National Aeronautics and Space Administration's Earth Science Technology Office,
Computation Technologies Project, under Cooperative Agreement Number NCC5-626 between
NASA and the California Institute of Technology, and \textsc{mpfit}, a Python script
that uses the Levenberg-Marquardt technique \citep{More1978} to solve least-squares
problems, based on an original \textsc{fortran} code part of the \textsc{minpack}-1 package. 

This research has also made use of the \textsc{aladin} interactive sky atlas \citep{Bonnarel2000}, of \textsc{saoimage ds9} \citep{Joye2003} developed by Smithsonian Astrophysical Observatory, of NASA's Astrophysics Data System, and of the NASA/IPAC Extragalactic Database \citep[NED;][]{Helou1991} which is operated by the Jet Propulsion Laboratory, California Institute of Technology, under contract with the National Aeronautics and Space Administration. 
 
Based on observations made with ESO Telescopes at the La Silla Paranal Observatory under programme ID 60.A-9317[A]. This work was co-funded under the Marie Curie Actions of the European Commission (FP7-COFUND). EP acknowledges support from the Spanish MINECO through project AYA2014-57490-P. LVM acknowledges support from AYA2015-65973-C3-1-R and AYA2014-52013-C2-1-R grants (MINECO Spain/FEDER, UE).

\end{acknowledgements}

\bibliographystyle{aa} 
\bibliography{bibliography_fixed} 

\begin{thebibliography}{70}
\expandafter\ifx\csname natexlab\endcsname\relax\def\natexlab#1{#1}\fi

\bibitem[{{Astropy Collaboration} {et~al.}(2013){Astropy Collaboration},
  Robitaille, Tollerud, Greenfield, Droettboom, Bray, Aldcroft, Davis,
  Ginsburg, Price-Whelan, Kerzendorf, Conley, Crighton, Barbary, Muna,
  Ferguson, Grollier, Parikh, Nair, Unther, Deil, Woillez, Conseil, Kramer,
  Turner, Singer, Fox, Weaver, Zabalza, Edwards, Azalee~Bostroem, Burke, Casey,
  Crawford, Dencheva, Ely, Jenness, Labrie, Lim, Pierfederici, Pontzen, Ptak,
  Refsdal, Servillat, \& Streicher}]{AstropyCollaboration2013}
{Astropy Collaboration}, Robitaille, T.~P., Tollerud, E.~J., {et~al.} 2013,
  \aap, 558, A33

\bibitem[{Baba {et~al.}(2015)Baba, Morokuma-Matsui, \& Egusa}]{Baba2015}
Baba, J., Morokuma-Matsui, K., \& Egusa, F. 2015, \pasj, 67, L4

\bibitem[{Berg {et~al.}(2015)Berg, Skillman, Croxall, Pogge, Moustakas, \&
  Johnson-Groh}]{Berg2015}
Berg, D.~A., Skillman, E.~D., Croxall, K.~V., {et~al.} 2015, \apj, 806, 16

\bibitem[{Bitsakis {et~al.}(2014)Bitsakis, Charmandaris, Appleton,
  D{\'\i}az-Santos, Le~Floc'h, {da Cunha}, Alatalo, \& Cluver}]{Bitsakis2014}
Bitsakis, T., Charmandaris, V., Appleton, P.~N., {et~al.} 2014, \aap, 565, A25

\bibitem[{Blanc {et~al.}(2013)Blanc, Weinzirl, Song, Heiderman, Gebhardt,
  Jogee, Evans, {van den Bosch}, Luo, Drory, Fabricius, Fisher, Hao, Kaplan,
  Marinova, Vutisalchavakul, \& Yoachim}]{Blanc2013}
Blanc, G.~A., Weinzirl, T., Song, M., {et~al.} 2013, \aj, 145, 138

\bibitem[{Bonnarel {et~al.}(2000)Bonnarel, Fernique, Bienaym{\'e}, Egret,
  Genova, Louys, Ochsenbein, Wenger, \& Bartlett}]{Bonnarel2000}
Bonnarel, F., Fernique, P., Bienaym{\'e}, O., {et~al.} 2000, \aap Supplement
  Series, 143, 33

\bibitem[{Borthakur {et~al.}(2010)Borthakur, Yun, \&
  Verdes-Montenegro}]{Borthakur2010}
Borthakur, S., Yun, M.~S., \& Verdes-Montenegro, L. 2010, \apj, 710, 385

\bibitem[{Bresolin(2011)}]{Bresolin2011}
Bresolin, F. 2011, \apj, 730, 129

\bibitem[{Bresolin {et~al.}(2009)Bresolin, Ryan-Weber, Kennicutt, \&
  Goddard}]{Bresolin2009}
Bresolin, F., Ryan-Weber, E., Kennicutt, R.~C., \& Goddard, Q. 2009, \apj, 695,
  580

\bibitem[{Calia {et~al.}(2014)Calia, Hackenberg, Holzl{\"o}hner, Lewis, \&
  Pfrommer}]{Calia2014}
Calia, D.~B., Hackenberg, W., Holzl{\"o}hner, R., Lewis, S., \& Pfrommer, T.
  2014, Advanced Optical Technologies, 3, 345

\bibitem[{Calzetti {et~al.}(2000)Calzetti, Armus, Bohlin, Kinney, Koornneef, \&
  Storchi-Bergmann}]{Calzetti2000}
Calzetti, D., Armus, L., Bohlin, R.~C., {et~al.} 2000, \apj, 533, 682

\bibitem[{Cappellari \& Emsellem(2004)}]{Cappellari2004}
Cappellari, M. \& Emsellem, E. 2004, \pasp, 116, 138

\bibitem[{Childress {et~al.}(2014{\natexlab{a}})Childress, Vogt, Nielsen, \&
  Sharp}]{Childress2014a}
Childress, M., Vogt, F., Nielsen, J., \& Sharp, R. 2014{\natexlab{a}},
  Astrophysics Source Code Library, ascl:1402.034

\bibitem[{Childress {et~al.}(2014{\natexlab{b}})Childress, Vogt, Nielsen, \&
  Sharp}]{Childress2014}
Childress, M.~J., Vogt, F. P.~A., Nielsen, J., \& Sharp, R.~G.
  2014{\natexlab{b}}, \apss, 349, 617

\bibitem[{Cid~Fernandes {et~al.}(2013)Cid~Fernandes, P{\'e}rez,
  Garc{\'\i}a~Benito, Gonz{\'a}lez~Delgado, {de Amorim}, S{\'a}nchez, Husemann,
  Falc{\'o}n~Barroso, S{\'a}nchez-Bl{\'a}zquez, Walcher, \&
  Mast}]{CidFernandes2013}
Cid~Fernandes, R., P{\'e}rez, E., Garc{\'\i}a~Benito, R., {et~al.} 2013, \aap,
  557, A86

\bibitem[{Cleveland(1979)}]{Cleveland1979}
Cleveland, W.~S. 1979, Journal of the American Statistical Association, 74, 829

\bibitem[{Croxall {et~al.}(2016)Croxall, Pogge, Berg, Skillman, \&
  Moustakas}]{Croxall2016}
Croxall, K., Pogge, R.~W., Berg, D.~A., Skillman, E.~D., \& Moustakas, J. 2016,
  ArXiv e-prints, 1605, arXiv:1605.01612

\bibitem[{{de Vaucouleurs} {et~al.}(1991){de Vaucouleurs}, {de Vaucouleurs},
  Corwin, Buta, Paturel, \& Fouqu{\'e}}]{deVaucouleurs1991}
{de Vaucouleurs}, G., {de Vaucouleurs}, A., Corwin, Jr., H.~G., {et~al.} 1991,
  Third {{Reference Catalogue}} of {{Bright Galaxies}}. {{Volume I}}:
  {{Explanations}} and References. {{Volume II}}: {{Data}} for Galaxies between
  0h and 12h. {{Volume III}}: {{Data}} for Galaxies between 12h and 24h.

\bibitem[{Dobbs \& Baba(2014)}]{Dobbs2014}
Dobbs, C. \& Baba, J. 2014, Publications of the Astronomical Society of
  Australia, 31, e035

\bibitem[{Dopita {et~al.}(2007)Dopita, Hart, McGregor, Oates, Bloxham, \&
  Jones}]{Dopita2007}
Dopita, M., Hart, J., McGregor, P., {et~al.} 2007, \apss, 310, 255

\bibitem[{Dopita {et~al.}(2010)Dopita, Rhee, Farage, McGregor, Bloxham, Green,
  Roberts, Neilson, Wilson, Young, Firth, Busarello, \& Merluzzi}]{Dopita2010}
Dopita, M., Rhee, J., Farage, C., {et~al.} 2010, \apss, 327, 245

\bibitem[{Dopita {et~al.}(2013)Dopita, Sutherland, Nicholls, Kewley, \&
  Vogt}]{Dopita2013}
Dopita, M.~A., Sutherland, R.~S., Nicholls, D.~C., Kewley, L.~J., \& Vogt, F.
  P.~A. 2013, \apjs, 208, 10

\bibitem[{Falc{\'o}n-Barroso {et~al.}(2011)Falc{\'o}n-Barroso,
  S{\'a}nchez-Bl{\'a}zquez, Vazdekis, Ricciardelli, Cardiel, Cenarro, Gorgas,
  \& Peletier}]{Falcon-Barroso2011}
Falc{\'o}n-Barroso, J., S{\'a}nchez-Bl{\'a}zquez, P., Vazdekis, A., {et~al.}
  2011, \aap, 532, A95

\bibitem[{Fischera \& Dopita(2005)}]{Fischera2005}
Fischera, J. \& Dopita, M. 2005, \apj, 619, 340

\bibitem[{Fitzpatrick(1999)}]{Fitzpatrick1999}
Fitzpatrick, E.~L. 1999, \pasp, 111, 63

\bibitem[{Freudling {et~al.}(2013)Freudling, Romaniello, Bramich, Ballester,
  Forchi, Garc{\'\i}a-Dabl{\'o}, Moehler, \& Neeser}]{Freudling2013}
Freudling, W., Romaniello, M., Bramich, D.~M., {et~al.} 2013, \aap, 559, A96

\bibitem[{Girardi {et~al.}(2000)Girardi, Bressan, Bertelli, \&
  Chiosi}]{Girardi2000}
Girardi, L., Bressan, A., Bertelli, G., \& Chiosi, C. 2000, \aap Supplement
  Series, 141, 371

\bibitem[{Grand {et~al.}(2012)Grand, Kawata, \& Cropper}]{Grand2012}
Grand, R. J.~J., Kawata, D., \& Cropper, M. 2012, \mnras, 421, 1529

\bibitem[{Grand {et~al.}(2016)Grand, Springel, Kawata, Minchev,
  S{\'a}nchez-Bl{\'a}zquez, G{\'o}mez, Marinacci, Pakmor, \&
  Campbell}]{Grand2016}
Grand, R. J.~J., Springel, V., Kawata, D., {et~al.} 2016, ArXiv e-prints, 1604,
  arXiv:1604.01027

\bibitem[{Helou {et~al.}(1991)Helou, Madore, Schmitz, Bicay, Wu, \&
  Bennett}]{Helou1991}
Helou, G., Madore, B.~F., Schmitz, M., {et~al.} 1991, in , 89--106

\bibitem[{Ho {et~al.}(2016)Ho, Medling, Groves, Rich, Rupke, Hampton, Kewley,
  Bland-Hawthorn, Croom, Richards, Schaefer, Sharp, \& Sweet}]{Ho2016}
Ho, I.-T., Medling, A.~M., Groves, B., {et~al.} 2016, \apss, 361, \#280

\bibitem[{Hunter(2007)}]{Hunter2007}
Hunter, J.~D. 2007, Computing in Science and Engineering, 9, 90

\bibitem[{Joye \& Mandel(2003)}]{Joye2003}
Joye, W.~A. \& Mandel, E. 2003, in , 489

\bibitem[{Kaplan {et~al.}(2016)Kaplan, Jogee, Kewley, Blanc, Weinzirl, Song,
  Drory, Luo, \& {van den Bosch}}]{Kaplan2016}
Kaplan, K.~F., Jogee, S., Kewley, L., {et~al.} 2016, \mnras, 462, 1642

\bibitem[{Kennicutt \& Garnett(1996)}]{Kennicutt1996}
Kennicutt, Jr., R.~C. \& Garnett, D.~R. 1996, \apj, 456, 504

\bibitem[{Kewley \& Ellison(2008)}]{Kewley2008}
Kewley, L.~J. \& Ellison, S.~L. 2008, \apj, 681, 1183

\bibitem[{Krajnovi{\'c} {et~al.}(2006)Krajnovi{\'c}, Cappellari, {de Zeeuw}, \&
  Copin}]{Krajnovic2006}
Krajnovi{\'c}, D., Cappellari, M., {de Zeeuw}, P.~T., \& Copin, Y. 2006,
  \mnras, 366, 787

\bibitem[{Kreckel {et~al.}(2016)Kreckel, Blanc, Schinnerer, Groves, Adamo,
  Hughes, \& Meidt}]{Kreckel2016}
Kreckel, K., Blanc, G.~A., Schinnerer, E., {et~al.} 2016, \apj, 827, 103

\bibitem[{Kroupa(2001)}]{Kroupa2001}
Kroupa, P. 2001, \mnras, 322, 231

\bibitem[{Li {et~al.}(2013)Li, Bresolin, \& Kennicutt}]{Li2013}
Li, Y., Bresolin, F., \& Kennicutt, Jr., R.~C. 2013, \apj, 766, 17

\bibitem[{L{\'o}pez-S{\'a}nchez {et~al.}(2012)L{\'o}pez-S{\'a}nchez, Dopita,
  Kewley, Zahid, Nicholls, \& Scharw{\"a}chter}]{Lopez-Sanchez2012}
L{\'o}pez-S{\'a}nchez, {\'A}.~R., Dopita, M.~A., Kewley, L.~J., {et~al.} 2012,
  \mnras, 426, 2630

\bibitem[{Luridiana {et~al.}(2013)Luridiana, Morisset, \& Shaw}]{Luridiana2013}
Luridiana, V., Morisset, C., \& Shaw, R.~A. 2013, Astrophysics Source Code
  Library, ascl:1304.021

\bibitem[{Luridiana {et~al.}(2015)Luridiana, Morisset, \& Shaw}]{Luridiana2015}
Luridiana, V., Morisset, C., \& Shaw, R.~A. 2015, \aap, 573, A42

\bibitem[{Marino {et~al.}(2013)Marino, Rosales-Ortega, S{\'a}nchez, {Gil de
  Paz}, V{\'\i}lchez, Miralles-Caballero, Kehrig, P{\'e}rez-Montero, Stanishev,
  Iglesias-P{\'a}ramo, D{\'\i}az, Castillo-Morales, Kennicutt,
  L{\'o}pez-S{\'a}nchez, Galbany, Garc{\'\i}a-Benito, Mast, Mendez-Abreu,
  Monreal-Ibero, Husemann, Walcher, Garc{\'\i}a-Lorenzo, Masegosa, Del
  Olmo~Orozco, Mour{\~a}o, Ziegler, Moll{\'a}, Papaderos,
  S{\'a}nchez-Bl{\'a}zquez, Gonz{\'a}lez~Delgado, Falc{\'o}n-Barroso, Roth,
  {van de Ven}, \& {Califa Team}}]{Marino2013}
Marino, R.~A., Rosales-Ortega, F.~F., S{\'a}nchez, S.~F., {et~al.} 2013, \aap,
  559, A114

\bibitem[{Mathewson {et~al.}(2013)Mathewson, Hart, Wehner, Hovey, \& {van
  Harmelen}}]{Mathewson2013}
Mathewson, D.~S., Hart, J., Wehner, H.~P., Hovey, G.~R., \& {van Harmelen}, J.
  2013, Journal of Astronomical History and Heritage, 16, 2

\bibitem[{Mor{\'e}(1978)}]{More1978}
Mor{\'e}, J.~J. 1978, in Numerical {{Analysis}}: {{Proceedings}} of the
  {{Biennial Conference Held}} at {{Dundee}}, {{June}} 28--{{July}} 1, 1977
  ({Springer Berlin Heidelberg}), 105--116

\bibitem[{Nicholls {et~al.}(2012)Nicholls, Dopita, \&
  Sutherland}]{Nicholls2012}
Nicholls, D.~C., Dopita, M.~A., \& Sutherland, R.~S. 2012, \apj, 752, 148

\bibitem[{Nicholls {et~al.}(2013)Nicholls, Dopita, Sutherland, Kewley, \&
  Palay}]{Nicholls2013}
Nicholls, D.~C., Dopita, M.~A., Sutherland, R.~S., Kewley, L.~J., \& Palay, E.
  2013, \apjs, 207, 21

\bibitem[{Rosales-Ortega {et~al.}(2011)Rosales-Ortega, D{\'\i}az, Kennicutt, \&
  S{\'a}nchez}]{Rosales-Ortega2011}
Rosales-Ortega, F.~F., D{\'\i}az, A.~I., Kennicutt, R.~C., \& S{\'a}nchez,
  S.~F. 2011, \mnras, 415, 2439

\bibitem[{Rosales-Ortega {et~al.}(2010)Rosales-Ortega, Kennicutt, S{\'a}nchez,
  D{\'\i}az, Pasquali, Johnson, \& Hao}]{Rosales-Ortega2010}
Rosales-Ortega, F.~F., Kennicutt, R.~C., S{\'a}nchez, S.~F., {et~al.} 2010,
  \mnras, 405, 735

\bibitem[{Rosolowsky \& Simon(2008)}]{Rosolowsky2008}
Rosolowsky, E. \& Simon, J.~D. 2008, \apj, 675, 1213

\bibitem[{S{\'a}nchez {et~al.}(2015)S{\'a}nchez, Galbany, P{\'e}rez,
  S{\'a}nchez-Bl{\'a}zquez, Falc{\'o}n-Barroso, Rosales-Ortega,
  S{\'a}nchez-Menguiano, Marino, Kuncarayakti, Anderson, Kruehler,
  Cano-D{\'\i}az, Barrera-Ballesteros, \&
  Gonz{\'a}lez-Gonz{\'a}lez}]{Sanchez2015}
S{\'a}nchez, S.~F., Galbany, L., P{\'e}rez, E., {et~al.} 2015, \aap, 573, A105

\bibitem[{S{\'a}nchez {et~al.}(2012{\natexlab{a}})S{\'a}nchez, Kennicutt, {Gil
  de Paz}, {van de Ven}, V{\'\i}lchez, Wisotzki, Walcher, Mast, Aguerri,
  Albiol-P{\'e}rez, Alonso-Herrero, Alves, Bakos, Bart{\'a}kov{\'a},
  Bland-Hawthorn, Boselli, Bomans, Castillo-Morales, Cortijo-Ferrero, {de
  Lorenzo-C{\'a}ceres}, Del~Olmo, Dettmar, D{\'\i}az, Ellis,
  Falc{\'o}n-Barroso, Flores, Gallazzi, Garc{\'\i}a-Lorenzo,
  Gonz{\'a}lez~Delgado, Gruel, Haines, Hao, Husemann, Igl{\'e}sias-P{\'a}ramo,
  Jahnke, Johnson, Jungwiert, Kalinova, Kehrig, Kupko, L{\'o}pez-S{\'a}nchez,
  Lyubenova, Marino, M{\'a}rmol-Queralt{\'o}, M{\'a}rquez, Masegosa, Meidt,
  Mendez-Abreu, Monreal-Ibero, Montijo, Mour{\~a}o, Palacios-Navarro,
  Papaderos, Pasquali, Peletier, P{\'e}rez, P{\'e}rez, Quirrenbach, Rela{\~n}o,
  Rosales-Ortega, Roth, Ruiz-Lara, S{\'a}nchez-Bl{\'a}zquez, Sengupta, Singh,
  Stanishev, Trager, Vazdekis, Viironen, Wild, Zibetti, \&
  Ziegler}]{Sanchez2012a}
S{\'a}nchez, S.~F., Kennicutt, R.~C., {Gil de Paz}, A., {et~al.}
  2012{\natexlab{a}}, \aap, 538, A8

\bibitem[{S{\'a}nchez {et~al.}(2016{\natexlab{a}})S{\'a}nchez, P{\'e}rez,
  S{\'a}nchez-Bl{\'a}zquez, Garc{\'\i}a-Benito, Ibarra-Mede, Gonz{\'a}lez,
  Rosales-Ortega, S{\'a}nchez-Menguiano, Ascasibar, Bitsakis, Law,
  Cano-D{\'\i}az, L{\'o}pez-Cob{\'a}, Marino, {Gil de Paz},
  L{\'o}pez-S{\'a}nchez, Barrera-Ballesteros, Galbany, Mast, Abril-Melgarejo,
  \& Roman-Lopes}]{Sanchez2016a}
S{\'a}nchez, S.~F., P{\'e}rez, E., S{\'a}nchez-Bl{\'a}zquez, P., {et~al.}
  2016{\natexlab{a}}, \rmxaa, 52, 171

\bibitem[{S{\'a}nchez {et~al.}(2016{\natexlab{b}})S{\'a}nchez, P{\'e}rez,
  S{\'a}nchez-Bl{\'a}zquez, Gonz{\'a}lez, Ros{\'a}lez-Ortega, {Cano-D{\'\i}
  az}, L{\'o}pez-Cob{\'a}, Marino, {Gil de Paz}, Moll{\'a},
  L{\'o}pez-S{\'a}nchez, Ascasibar, \& Barrera-Ballesteros}]{Sanchez2016}
S{\'a}nchez, S.~F., P{\'e}rez, E., S{\'a}nchez-Bl{\'a}zquez, P., {et~al.}
  2016{\natexlab{b}}, \rmxaa, 52, 21

\bibitem[{S{\'a}nchez {et~al.}(2014)S{\'a}nchez, Rosales-Ortega,
  Iglesias-P{\'a}ramo, Moll{\'a}, Barrera-Ballesteros, Marino, P{\'e}rez,
  S{\'a}nchez-Blazquez, Gonz{\'a}lez~Delgado, Cid~Fernandes, {de
  Lorenzo-C{\'a}ceres}, Mendez-Abreu, Galbany, Falcon-Barroso,
  Miralles-Caballero, Husemann, Garc{\'\i}a-Benito, Mast, Walcher, {Gil de
  Paz}, Garc{\'\i}a-Lorenzo, Jungwiert, V{\'\i}lchez, J{\'\i}lkov{\'a},
  Lyubenova, Cortijo-Ferrero, D{\'\i}az, Wisotzki, M{\'a}rquez, Bland-Hawthorn,
  Ellis, {van de Ven}, Jahnke, Papaderos, Gomes, Mendoza, \&
  L{\'o}pez-S{\'a}nchez}]{Sanchez2014}
S{\'a}nchez, S.~F., Rosales-Ortega, F.~F., Iglesias-P{\'a}ramo, J., {et~al.}
  2014, \aap, 563, A49

\bibitem[{S{\'a}nchez {et~al.}(2012{\natexlab{b}})S{\'a}nchez, Rosales-Ortega,
  Marino, Iglesias-P{\'a}ramo, V{\'\i}lchez, Kennicutt, D{\'\i}az, Mast,
  Monreal-Ibero, Garc{\'\i}a-Benito, Bland-Hawthorn, P{\'e}rez,
  Gonz{\'a}lez~Delgado, Husemann, L{\'o}pez-S{\'a}nchez, Cid~Fernandes, Kehrig,
  Walcher, {Gil de Paz}, \& Ellis}]{Sanchez2012}
S{\'a}nchez, S.~F., Rosales-Ortega, F.~F., Marino, R.~A., {et~al.}
  2012{\natexlab{b}}, \aap, 546, A2

\bibitem[{S{\'a}nchez-Bl{\'a}zquez {et~al.}(2006)S{\'a}nchez-Bl{\'a}zquez,
  Peletier, Jim{\'e}nez-Vicente, Cardiel, Cenarro, Falc{\'o}n-Barroso, Gorgas,
  Selam, \& Vazdekis}]{Sanchez-Blazquez2006}
S{\'a}nchez-Bl{\'a}zquez, P., Peletier, R.~F., Jim{\'e}nez-Vicente, J.,
  {et~al.} 2006, \mnras, 371, 703

\bibitem[{S{\'a}nchez-Menguiano
  {et~al.}(2016{\natexlab{a}})S{\'a}nchez-Menguiano, S{\'a}nchez, Kawata,
  Chemin, P{\'e}rez, Ruiz-Lara, S{\'a}nchez-Bl{\'a}zquez, Galbany, Anderson,
  Grand, Minchev, \& G{\'o}mez}]{Sanchez-Menguiano2016a}
S{\'a}nchez-Menguiano, L., S{\'a}nchez, S.~F., Kawata, D., {et~al.}
  2016{\natexlab{a}}, ArXiv e-prints, 1610, arXiv:1610.00440

\bibitem[{S{\'a}nchez-Menguiano
  {et~al.}(2016{\natexlab{b}})S{\'a}nchez-Menguiano, S{\'a}nchez, P{\'e}rez,
  Garc{\'\i}a-Benito, Husemann, Mast, Mendoza, Ruiz-Lara, Ascasibar,
  Bland-Hawthorn, Cavichia, D{\'\i}az, Florido, Galbany, G{\'o}nzalez~Delgado,
  Kehrig, Marino, M{\'a}rquez, Masegosa, M{\'e}ndez-Abreu, Moll{\'a}, Del~Olmo,
  P{\'e}rez, S{\'a}nchez-Bl{\'a}zquez, Stanishev, Walcher,
  L{\'o}pez-S{\'a}nchez, \& {Califa Collaboration}}]{Sanchez-Menguiano2016}
S{\'a}nchez-Menguiano, L., S{\'a}nchez, S.~F., P{\'e}rez, I., {et~al.}
  2016{\natexlab{b}}, \aap, 587, A70

\bibitem[{Sanders {et~al.}(2012)Sanders, Caldwell, McDowell, \&
  Harding}]{Sanders2012}
Sanders, N.~E., Caldwell, N., McDowell, J., \& Harding, P. 2012, \apj, 758, 133

\bibitem[{Schlafly \& Finkbeiner(2011)}]{Schlafly2011}
Schlafly, E.~F. \& Finkbeiner, D.~P. 2011, \apj, 737, 103

\bibitem[{Schlegel {et~al.}(1998)Schlegel, Finkbeiner, \& Davis}]{Schlegel1998}
Schlegel, D.~J., Finkbeiner, D.~P., \& Davis, M. 1998, \apj, 500, 525

\bibitem[{Seabold \& Perktold(2010)}]{Seabold2010}
Seabold, S. \& Perktold, J. 2010, in Proc. of the 9th {{Python}} in {{Science
  Conference}}, 57--61

\bibitem[{Stuik {et~al.}(2006)Stuik, Bacon, Conzelmann, Delabre, Fedrigo,
  Hubin, Le~Louarn, \& Str{\"o}bele}]{Stuik2006}
Stuik, R., Bacon, R., Conzelmann, R., {et~al.} 2006, New Astronomy Reviews, 49,
  618

\bibitem[{Vazdekis {et~al.}(2010)Vazdekis, S{\'a}nchez-Bl{\'a}zquez,
  Falc{\'o}n-Barroso, Cenarro, Beasley, Cardiel, Gorgas, \&
  Peletier}]{Vazdekis2010}
Vazdekis, A., S{\'a}nchez-Bl{\'a}zquez, P., Falc{\'o}n-Barroso, J., {et~al.}
  2010, \mnras

\bibitem[{Vogt(2015)}]{Vogt-thesis}
Vogt, F. P.~A. 2015, PhD Thesis, Australian National University

\bibitem[{Vogt {et~al.}(2015)Vogt, Dopita, Borthakur, Verdes-Montenegro,
  Heckman, Yun, \& Chambers}]{Vogt2015}
Vogt, F. P.~A., Dopita, M.~A., Borthakur, S., {et~al.} 2015, \mnras, 450, 2593

\bibitem[{Vogt {et~al.}(2016)Vogt, Seitenzahl, Dopita, \& Ruiter}]{Vogt2016a}
Vogt, F. P.~A., Seitenzahl, I.~R., Dopita, M.~A., \& Ruiter, A.~J. 2016, ArXiv
  e-prints, 1611, arXiv:1611.03862

\bibitem[{Zinchenko {et~al.}(2016)Zinchenko, Pilyugin, Grebel, S{\'a}nchez, \&
  V{\'\i}lchez}]{Zinchenko2016}
Zinchenko, I.~A., Pilyugin, L.~S., Grebel, E.~K., S{\'a}nchez, S.~F., \&
  V{\'\i}lchez, J.~M. 2016, \mnras, 462, 2715

\end{thebibliography}

\appendix
\section{A brief description of \textsc{brutus}}\label{app:brutus}
We hereafter describe briefly the principal processing steps performed by \textsc{brutus} in the case of HCG~91c. The user interested in using this code should however bear in mind that the code is still under active development, and as such subject to evolve over time.

\textsc{brutus} is designed to require as little interaction as possible, in order to speed the processing of large (MUSE) datacubes. All parameters with scientific implications are fed by the user to the different routines using a dedicated input \textsc{pickled} file, thereby making the post-processing of a datacube largely \emph{hands-free}. \textsc{brutus} is able to exploit multiple CPUs to speed-up the processing of a given datacube, and aims at creating high-quality diagrams that fully exploit the world coordinate system (WCS) of the data. Other specificities of \textsc{brutus} include the facts that: 1) it is entirely written in \textsc{python}, with no outside software required (with the exception of \textsc{montage}), 2) it is able to cope with spectra with a non-uniform spectral resolution, as is the case for the MUSE instrument, 3) it allows both a \emph{spaxel-based} processing or an \emph{aperture-based} processing of a given datacube, with the in-built ability to identify structures in the data (e.g. H\,{\textsc\smaller II} regions) and create dedicated apertures, and 4) it exploits existing \textsc{python} tools with proven track-records developed by the community and individuals, including \textsc{astropy} \citep{AstropyCollaboration2013}, \textsc{statsmodels} \citep{Seabold2010}, \textsc{aplpy}, \textsc{ppxf} \citep{Cappellari2004} and \textsc{fit\_kinematics\_pa} \citep{Krajnovic2006}. 

Once again, we stress that \textsc{brutus} is a work in progress, and should not yet be considered complete or fully stable: existing modules are being updated and new ones added depending on the specific needs of our team, and so is the link to additional useful community packages \citep[e.g. \textsc{pyneb} is a foreseen addition;][]{Luridiana2013,Luridiana2015}. Contributions from the community that help make the code more polyvalent, more reliable and/or expand its capabilities are also welcome through the Github interface. Finally, we note that \textsc{brutus} is built to accommodate (in principle) data from any integral fields spectrograph (both in the visible or in the IR), but it has so far been developed and tested with MUSE datacubes only.

\subsection{Continuum and emission line fitting}

After performing an initial measure of the signal-to-noise ratio (SNR) in the continuum and H$\alpha$ emission line for all spaxels in the datacube, each spaxel is corrected for the Galactic extinction along the line of sight towards HCG~91c. The Galactic extinction is derived via the Nasa Extragalactic Database (NED) from the \cite{Schlafly2011} recalibration of the \cite{Schlegel1998} infrared-based dust map: for HCG~91c, $A_B=0.069$ and $A_V=0.052$. The map is based on dust emission from COBE/DIRBE and IRAS/ISSA. The recalibration assumes a \cite{Fitzpatrick1999} reddening law with $R_V = 3.1$ and a different source spectrum than \cite{Schlegel1998}.

Each spectrum with SNR$<$3 in the continuum is then fitted using the non-parametric Locally-Weighted Scatterplot Smoothing (LOWESS) technique \citep{Cleveland1979}, via the \textsc{nonparametric.smoothers\_lowess.lowess()} routine inside the \textsc{statsmodels} module. This technique is ideal for fitting the stellar continuum at low SNR when fitting stellar population models isn't reliable at all. The technique is robust to the presence of emission lines and/or bad pixels in the spectra, in that if can efficiently ignore them using a multiple-pass approach. We note that this technique ought to also be of interest for fitting the nebular continuum (as an alternative to parametric functions) for galactic targets such as planetary nebulae or H\,{\textsc{\smaller II}} regions \citep[see e.g.][for the case of a supernova remnant]{Vogt2016a}.

The stellar continuum for all spaxels with SNR$\geq$3 is fitted using the \textsc{ppxf} code \citep{Cappellari2004}. For such low SNR, great caution must be used when interpreting the scientific implications of the fit. In this article, we solely focus on the analysis of the strong emission lines in the datacube. Fitting the stellar continuum on a spaxel-by-spaxel basis allows us to derive a \emph{reasonable estimate} of the amount of stellar absorption below the H$\alpha$ and H$\beta$ emission lines in the data - absorption features which become visible at SNR$\cong$3. Here, we fed \textsc{ppxf} with stellar population synthesis model predictions from the MILES stellar libraries \citep{Sanchez-Blazquez2006, Falcon-Barroso2011} at FWHM=2.5\,\AA, based on the code presented in \cite{Vazdekis2010}, using a Kroupa revised IMF with slope 1.3 \citep{Kroupa2001}, Padova+00 isochrones \citep{Girardi2000}, with 6 metallicity steps ranging from -1.71 to +0.22, and 13 ages ranging from 0.0631 to 15.8489 Gyr. We experimented with different model parameters, and note that our final choice does not influence the results described in the article at any significant level.

Emission lines are fitted in a subsequent \textsc{brutus} processing step, after subtracting the stellar continuum fitted either using the LOWESS algorithm (for spaxels with SNR<3) or \textsc{ppxf} (for spaxels with SNR$\geq$3). We rely on the \textsc{mpfit} module that uses the Levenberg-Marquardt technique \citep{More1978} to solve least-squares problems, based on an original Fortran code part of the \textsc{minpack}-1 package. Specifically, we use the version of \textsc{mpfit} included by M. Cappellari inside the Python package files of \textsc{ppxf}, but with a dedicated \textsc{brutus} wrapper. The following strong emission lines are fitted simultaneously, all with the same (tied) velocity:
\begin{eqnarray}
G1&:&\mathrm{H}\beta, [\mathrm{O}\,\textsc{\smaller III}] \lambda4959, [\mathrm{O}\,\textsc{\smaller III}] \lambda 5007\\
G2&:&[\mathrm{N}\,\textsc{\smaller II}] \lambda 6548, \mathrm{H}\alpha, [\mathrm{N}\,\textsc{\smaller II}] \lambda 6583,\nonumber\\
&&[\mathrm{S}\,\textsc{\smaller II}] \lambda 6716,[\mathrm{S}\,\textsc{\smaller II}] \lambda6731.
\end{eqnarray}

The velocity dispersion was tied for all $G1$ and $G2$ lines respectively, as a consequence of the varying spectral resolution of MUSE. Whereas in principle \textsc{brutus} can allow for all lines to be fitted with individual velocity dispersions, the time-cost is (at the time of submission of this article) prohibitive in practice, and the fit less reliable for low SNR spectra. A single component fit is appropriate for all spaxels, as anticipated from the observations of HCG~91c acquired with a spectral resolution of $R=7000$ using the WiFeS integral field spectrograph \citep{Dopita2007,Dopita2010} as discussed in \cite{Vogt2015}. 

\subsection{Semi-automated detection of H\,\textsc{\smaller II} regions}

\textsc{brutus} relies on the \textsc{find\_peaks()} function inside the \textsc{photutils} module to identify local maxima throughout the H$\alpha$ emission flux map. To each local maxima, \textsc{brutus} assigns a fixed-size aperture, with a radius set to 0.6\,arcsec$=$3 spaxels. For most H\,\textsc{\smaller II} regions in HCG~91c, a fixed-size aperture is a suitable choice as the star-forming complexes are clearly distinguishable, but not individually resolved. Nonetheless, \textsc{brutus} also offers the ability to interactively add or remove apertures of varying radii via an interactive \textsc{matplotlib} window. The ability to rapidly add or remove additional apertures with a simple mouse click allows to rapidly and efficiently refine the automated selection -necessarily imperfect- by removing spurious detections, adding missing apertures, and adjusting the aperture sizes for extended star-forming complexes. At the time of submission, this interactive inspection of the aperture list is the only (optional) manual step within \textsc{brutus}. We stress that the \textit{interactivity} of this step is a strong factor allowing to speed-up the procedure: for the case of HCG~91c, revising (visually) the list of 556 H\,\textsc{\smaller II} regions can be done in a matter of a few minutes.

From the list of apertures (i.e. their center and radius), a new data cube is generated, where each spaxel within a given aperture is replaced with the sum of all the spaxels inside the said aperture. All other spaxels (not associated to any aperture) are left empty (i.e. as \textsc{python}'s \textsc{nan}s). This results in a (somewhat redundant) cube that can be fed back into \textsc{brutus}, and processed exactly as the normal cube - but with a crucial improvement in the SNR associated with the fainter H\,\textsc{\smaller II} regions in the outer regions of the galaxy.

\end{document}